\documentclass[journal]{IEEEtran}
\usepackage{amsfonts}
\usepackage{amssymb}
\usepackage{eurosym}
\usepackage{cite}
\usepackage{graphicx}
\usepackage{epstopdf}
\usepackage{amsmath}
\usepackage{tikz,lipsum}
\usepackage{caption}
\usepackage[T1]{fontenc}
\usepackage{amsthm}
\usepackage{mathrsfs}
\usepackage{color}
\usepackage{stfloats}
\usepackage{xcolor, etoolbox}
\usepackage{subcaption}

\IEEEoverridecommandlockouts
\newcounter{mytempeqncnt}
\newtheorem{remark}{Remark}

\newtheorem{proposition}{Proposition}

\begin{document}

\title{On the Physical Layer Security of a Dual-Hop UAV-based Network in the Presence of per-hop Eavesdropping and Imperfect CSI}
\author{Elmehdi Illi, \IEEEmembership{Member, IEEE}, Marwa K. Qaraqe, %
\IEEEmembership{Senior Member, IEEE}, Faissal El Bouanani, %
\IEEEmembership{Senior Member, IEEE}, and Saif M. Al-Kuwari,
\IEEEmembership{Senior Member,
IEEE} \thanks{%
This research was sponsored in part by the NATO Science for Peace and Security Programme under grant SPS G5797.}
\thanks{This paper was submitted
in part to the IEEE Global Communications Conference
 (Globecom'22) \cite{gc2022}}
\thanks{%
E. Illi, M. K. Qaraqe, and S. M. Al-Kuwari are with the College of Science
and Engineering, Hamad Bin Khalifa University, Qatar Foundation, Doha,
Qatar. (e-mails: elmehdi.illi@ieee.org, \{mqaraqe, smalkuwari\}@hbku.edu.qa).%
} \thanks{%
F. El Bouanani is with ENSIAS College of Engineering, Mohammed V University
of Rabat, Morocco. (e-mail: f.elbouanani@um5s.net.ma).} }
\maketitle

\begin{abstract}
In this paper, the physical layer security of a dual-hop unmanned aerial vehicle-based wireless network, subject to imperfect channel state information (CSI) and mobility effects, is analyzed. Specifically, a source node $(S)$ communicates with a destination node $(D)$ through a decode-and-forward relay $(R)$, in the presence of two wiretappers $\left(E_{1},E_{2}\right)$ independently trying to compromise the two hops. Furthermore, the transmit nodes $\left(S,R\right) $ have a single transmit antenna, while the receivers $\left(R,D,E_{1},E_{2}\right) $ are equipped with multiple receive antennas. Based on the per-hop signal-to-noise ratios (SNRs) and correlated secrecy capacities' statistics, a closed-form expression for the secrecy intercept probability (IP) metric is derived, in terms of key system parameters. Additionally, asymptotic expressions are
revealed for two scenarios, namely (i) mobile nodes with
imperfect CSI and (ii) static nodes with perfect CSI. The results show that a zero secrecy diversity order is manifested for the first scenario, due to the presence of a ceiling value of the average SNR, while the IP drops linearly at high average SNR in the second one, where the achievable diversity order depends on the fading parameters and number of antennas of the legitimate links/nodes. Furthermore, for static nodes, the system can be castigated by a $15$ dB secrecy loss at IP$=3\times10^{-3},$ when the CSI imperfection power raises from $0$ to $10^{-3}$. Lastly, the higher the legitimate nodes' speed, carrier frequency, delay, and/or relay's decoding threshold SNR, the worse is the system's secrecy. Monte Carlo simulations
endorse the derived analytical results.
\end{abstract}



\IEEEpeerreviewmaketitle

\section{Introduction}

With the rapid growth of wireless communication technologies throughout
the last few years, the Internet of Things (IoT) paradigm strongly emerged
as a leading actor of the fifth generation (5G) wireless network standard and beyond \cite%
{iot1}. Practically, IoT architectures and protocols can be implemented in
numerous promising services and applications such as traffic control, smart
home, smart grids, factories, and healthcare facilities \cite{ref1,ref2}.
Recent studies forecast around 500 billion connected IoT devices over
the globe by 2030 \cite{ref3}. Nonetheless, despite the aforesaid features
and applications, guaranteeing a fair functioning of such IoT\ networks is
challenging in underserved areas (e.g., deserts, mountains), or in disaster
zones non-covered by wireless cellular infrastructure \cite{ref4,ref1}.

Unmanned aerial vehicles\ (UAVs)\ have been attracting remarkable interest
on the wireless community as an alternative mean for real-time
communication and extended network coverage \cite{ref6}. UAVs
can be conveniently implemented in disaster regions with limited cellular
coverage to act as flying base stations (BSs), where they can provide
enhancements in both radio access and backhaul connectivity with
clearly improved coverage \cite{slimsurvey}. Additionally, due to
the high probability of the presence of line-of-sight links, UAVs\ can be involved
in other use-cases such as media production, real-time surveillance,
and automatic target detection \cite{uav1,uav3}.

Small UAVs (i.e., below 5Kg) generally operate in swarms \cite{uav1,uav2}.
In such a scenario, multihop relaying is implemented to reliably convey
information signals between a distant source and destination nodes via
routing techniques \cite{uavsurvey}. At the physical layer (PHY), two main relaying protocols are usually deployed, namely (i)
amplify-and-forward (AF), a.k.a non-regenerative relaying scheme, and (ii)
decode-and-forward (DF), a.k.a the regenerative relaying one. The former, although
less complex, can provide acceptable performance in noise-limited scenarios.
However, the higher the noise power, the worse its performance becomes due to noise amplification \cite{relaying}. On the other hand, DF
generally provides better performance, but
requires complex digital signal processing and computing circuits onboard
compared to the AF scheme \cite{relaying2}.

PHY\ security (PLS) has gained attention in the wireless communication community
from both academia and industry \cite{sustainable}. The open
broadcast nature of radio links exposes the legitimate communication to
permanent threats from malign users \cite{pla}. Therefore,
information security has been regarded as a fundamental requirement, particularly
on the 5GB\ and especially for IoT. Upper layers look at ensuring confidentiality and authentication
through key-based cryptographic schemes \cite{access}. Nonetheless, such
schemes are often faced by several challenges and limitations in IoT
networks, such as (i)\ the complexity of key generation and distribution in
large-scale decentralized networks, and (ii) the operational complexity of
advanced cryptographic schemes, which poses stringent burdens on low-power
devices \cite{jehad}. To this end, PLS
\footnote{PLS\ generally refers to the use of PHY\
parameters to ensure either data confidentiality and/or authentication. PLS can be broadly categorized as: PHY authentication (PLA) \cite{pla}, PHY\ key generation (PLKG)
\cite{plkg}, or conventional Wyner-inspired PLS. In this paper, we adopt the latter,
aiming at ensuring data confidentiality based on the system secrecy capacity.%
} has been a popular candidate for effective low-complexity security
solutions. Its core advantage lies in establishing key-less
information-theoretically secure communications by exploiting the random
nature of the propagation channel along with PHY\ parameters, to provide
high security levels with low overhead. Despite these advantages, PLS\ can
exhibit several challenges in UAV/vehicular networks, especially when the communication
is under the joint effect of imperfect channel estimation and nodes
mobility, leading to outdated channel state information (CSI) with
estimation errors \cite{anshul}.

\subsection{Related Work}

Throughout the literature, extensive research work has inspected the PLS of
dual-hop or multi-hop communication systems \cite%
{lit1,lit2,lit3,lit4,lit5,lit6,lit7,lit8,lit9,lit10,lit11,lit12,access}. The
authors in \cite{lit1} proposed relay selection schemes in a multiuser
network, considering AF\ relays to improve the secrecy level of the system.
In \cite{lit2,lit3}, the secrecy performance was investigated under the
joint use of maximal-ratio combining (MRC) and zero-forcing beamforming in a
cognitive-radio network (CRN) and a non-cognitive one, respectively. The
work in \cite{lit4} inspected secrecy metrics for a dual-hop DF-based
multiuser network, under the presence of various eavesdroppers overhearing
the second hop. Importantly, the authors of \cite%
{access,sustainable,lit6,lit8,lit9,lit11}, and \cite{mounia} analyzed the
end-to-end PLS\ of mixed radio-frequency (RF)-optical wireless communication
(OWC) systems, where the optical link operates either in indoor (i.e.,
visible light communications), outdoor (i.e., free-space optics), or in the
marine medium (i.e., underwater OWC).

Nonetheless, the above-mentioned work were restricted to (i) assuming the
presence of eavesdroppers in only one of the two hops, and (ii)\ the fact
that all nodes are static with perfect CSI. In fact, due to the nodes'
mobility, the channel between each pair of transceivers within the network
undergoes time selectivity, where the fading realizations decorrelate in
time according to the well-known Jakes' model \cite{rosa,goldsmith}. To this
end, the estimated CSI\ at the packet preamble will be outdated compared to
the actual one when processing the received signal. Furthermore, due to the
imperfections in the receivers' tracking loops, the CSI\ estimation suffers
from estimation noise in addition to the time selectivity \cite{khattabi}.

A number of other work from the literature tackled the secrecy performance of
dual-hop networks by encompassing the joint distortions due to the CSI\ time
selectivity and/or estimation errors. For instance, the authors in \cite{lit5}
investigated the secrecy performance of a dual-hop network with multi-antenna
nodes employing transmit antenna selection along with MRC, under the
presence of a single eavesdropper, and considering Nakagami-$m$ fading
channels. A similar setup was analyzed in \cite{lit16} by
assuming single-antenna nodes and a Rayleigh fading model. The work in \cite{lit12} dealt with the performance of an AF-based dual-hop scheme with
single-antenna mobile nodes and a single eavesdropper, where the two hops
are subject to Rayleigh and double Rayleigh fading models. The latter work
was extended in \cite{lit13} by considering the generalized Nakagami-$m$ and
double Nakagami-$m$ fading models. Moreover, \cite{lit14} investigated the
secrecy outage probability performance of a dual-hop mixed RF-FSO\ system,
where the CSI\ is subject to time selectivity and estimation errors. The
authors in \cite{lit15} analyzed the system's secrecy in a multi-relay
dual-hop scheme subject to the presence of one malicious node and assuming
single-antenna transceivers. In addition, the authors in \cite{lit17}
tackled the secrecy analysis of a dual-hop AF-based CRN, subject to Rayleigh
and double-Rayleigh fading and assuming single-antenna devices. In \cite%
{lit18}, the secrecy performance of a dual-hop CRN is carried out under
mobility constraints and assuming a single eavesdropper and single-antenna
devices. The performance of a relay-based
device-to-device network with full-duplex consideration and multiple-antenna
nodes was investigated in \cite{lit19} by considering yet again a single
eavesdropper. In \cite{lit20}, the
authors investigated the secrecy level of a dual-hop network operating with
various opportunistic relays and a single eavesdropper by considering a
correlation between the legitimate and wiretap channels. Similarly, the
authors carried out a secrecy performance analysis in \cite{lit21} for a
two-way dual-hop communication system by assuming a multi-antenna source and
destination nodes, along with a single malign one. Moreover, the
work in \cite{lit22,lit23} evaluated the security gains of employing
artificial noise/jamming techniques in cooperative dual-hop networks.

\subsection{Motivation}

Although all the above-mentioned work have inspected the secrecy of
cooperative dual-hop networks with the joint effect of imperfect CSI and
node mobility, all of these work assume either the presence of (i) a single
or multiple collaborative eavesdroppers within one hop, (ii) single-antenna
nodes, or (iii) the use of the basic Rayleigh fading model. In fact, few
work in the literature were reported to deal with independent eavesdroppers
attacking each hop. For instance, the authors in \cite{lit11} treated a
dual-hop FSO-RF\ scheme with two parallel paths, with the presence of
multiple eavesdroppers looking to compromise the RF hop. Likewise, the
authors in \cite{lit24} investigated the impact of fading and
turbulence on the secrecy performance of a dual-hop FSO-RF\ system with two
eavesdroppers attempting to independently intercept the FSO and RF\ hops.
A parallel analysis of a similar setup was
conducted in \cite{lit25} by adding the residual hardware impairments
constraint along with energy harvesting into the analysis. The authors in
\cite{ojcoms} analyzed the secrecy performance of a dual-hop
hybrid-terrestrial satellite (HTS) system with an optical feeder, subject to
the presence of an eavesdropper per each hop. Lastly, a similar setup was
analyzed in \cite{mounia} by considering an RF-FSO\ HTS\ system.
Nevertheless, the analysis in all these above-mentioned work, (\cite%
{lit24,lit25,lit11,ojcoms,mounia}), was restricted to a single-antenna
assumption in some of or all the nodes along with assuming static nodes with
perfect CSI estimation. At the same time, the presence of UAVs and vehicular networks
has been emphatically emerging in contemporary and futuristic networks.
In particular, flying base stations (e.g., UAVs, high-altitude platforms, etc.) are strongly advocated to be a key actor in space-to-ground
integrated communications, whereby the wireless communication community aims
at providing connectivity solutions to underserved or unconnected zones \cite%
{slimfrontiers}. Such mobile networks, often organized in swarms \cite{uav1}, are prone to numerous threats and security challenges, in which potential
malign users attempt to independently compromise each hop of the communication
network. Therefore, it is crucial to conduct a secrecy analysis of such mobile networks, by
assuming CSI imperfections, mobility effects, and the presence of
independent per-hop eavesdroppers.

\subsection{Contributions}

Capitalizing on the above motivations, we aim in this paper to analyze the
PLS of a dual-hop UAV-assisted wireless communication
system (WCS) under the impact of node mobility and imperfect channel
estimation. In particular, a source node $\left( S\right) $ communicates
with a destination node $\left( D\right) $ through a DF-based relay $%
\left( R\right) $. At the same time, two eavesdroppers ($E_{1}$ and $E_{2}$) are
attempting to independently intercept the $S$-$R$ and $R$-$D$ communications,
respectively. It is assumed that all transmitters ($S$, $R$)
have single transmit antennas, while receivers ($R$, $D$, $E_{1}$, $E_{2}$) employ the MRC technique leveraging the multiple receive
antennas onboard. Moreover, all the channels are assumed to follow a Nakagami-$%
m$ fading model \footnote{
In addition to the Rician model, the Nakagami-$m$ fading model was proved in
several fields measurements to give good agreements with the fading envelope
distribution in UAV air-to-air channel measurements, operating in moderate
altitudes or open spaces \cite{goddemeier,uavsurveychannel}.}. Hence, this
work differs from \cite%
{lit5,lit16,lit12,lit13,lit14,lit15,lit17,lit18,lit19,lit20,lit21,lit22,lit23}
by considering independent eavesdroppers per each of the two hops, and
from \cite{lit24,lit25,lit11,ojcoms,mounia} by taking into account the
distortions due to mobility and CSI\ imperfect estimation. To the best of
our knowledge, our work is the first that takes into consideration
the joint effect of transceivers mobility and CSI estimation errors,
independent eavesdropping per each of the two hops, Nakagami-$m$ fading
model, and multi-antenna nodes. The key contributions of this work can be
summarized as follows:
\begin{itemize}
\item Leveraging the first-order autoregressive model along with the
Gaussian-distributed CSI estimation error, the instantaneous per-hop
signal-to-noise ratio (SNR) is expressed in terms of the mobility-dependent
correlation coefficient (function of the carrier frequency, relative speed,
and delay), transmit power, average fading and noise powers, CSI\ estimation
error variance, and the number of receive antennas.

\item Capitalizing on the statistical properties of the per-hop\ SNR, we derive an
exact closed-form expression for the system's intercept probability (IP)\
metric in terms of key system and channel parameters; namely, the
average per-branch SNR (function of the mobility-dependent correlation
coefficient, depending on the carrier frequency, relative speed, and delay), the
number of antennas onboard the legitimate and wiretap nodes, per-hop fading
severity parameter, and the relay decoding threshold SNR.

\item The impact of the main system parameters on the setup's security level
is broadly discussed based on the derived analytical expressions. In
particular, the IP\ behavior is inspected for particular setup parameters'
values, e.g., low and high values of decoding threshold SNR, various values
of fading severity parameters, and average fading powers.
\item Numerical and
simulation results are conducted to validate all the derived analytical
results. We show that the system's secrecy is
significantly influenced by nodes' relative speed, CSI estimation noise
level, threshold SNR fading severity parameters, and the number of antennas
onboard.

\item To obtain more insights and observations from the above results,
asymptotic expressions in the high SNR\ regime are provided for two
scenarios, namely (i)\ a generalized scenario of mobile nodes and imperfect
CSI, and (ii) an ideal scenario for static UAVs and perfect CSI\ estimation.
Based on these expressions, the respective secrecy diversity order is
retrieved for these two scenarios. We show that the secrecy
performance of the first scenario exhibits a zero diversity order at high
SNR, regardless of the number of antennas, fading parameters values, and
threshold SNR. Nevertheless, a secrecy diversity order of $\min \left(
m_{SR}N_{R},m_{RD}N_{D}\right) $ is reached for the second scenario, with $%
m_{SR}$ and $m_{RD}$ denote the fading severity of the $S$-$R$ and $R$-$D$
hops, respectively, while $N_{R}$ and $N_{D}$ refer to the number of
antennas onboard $R$ and $D$, respectively.
\end{itemize}

\subsection{Organization}

The remainder of this paper is organized as follows: Section II describes
the adopted system and channel models, and Section III provides useful
statistical properties for the per-hop SNRs. In Section IV, exact
closed-form and asymptotic expressions for the IP\ metric are provided with
several insights on the impact of key system parameters on the derived
results. Section V presents numerical results and discussions. Finally,
Section VI concludes the paper.

\subsection{Notations}

Vectors are denoted by bold letters, $\mathcal{CN}\left( \mu ,\sigma
^{2}\right) $ refers to complex Gaussian distribution with mean $\mu $ and
variance $\sigma ^{2},$ $\left\vert .\right\vert $ is the absolute value, $%
\mathbb{E}\left[ .\right] $ refers to the expected value of a random
variable, and $\overline{\mathbf{h}}$ and $\mathbf{\hat{h}}$ refer to the
outdated version and estimate of $\mathbf{h,}$ respectively$.$ Furthermore, $%
J_{n}\left( .\right) $~is the $n$th order modified Bessel function of the
first kind \cite[Eq (8.411.1)]{integrals} and $L^{2}$-Norm is denoted by $%
\left\Vert \mathbf{.}\right\Vert _{2}$. Additionally, $\Gamma \left(
.\right) $, $\gamma _{\text{inc}}\left( .,.\right) $, and $\Gamma _{\text{inc%
}}\left( .,.\right) $ indicate the complete, lower-incomplete, and
upper-incomplete Gamma functions, respectively \cite[Eqs (8.310.1, 8.350.1,
8.350.2)]{integrals}. Lastly, $_{2}F_{1}\left( .,.;.;.\right) $ is the Gauss
Hypergeometric function \cite[Eq. (07.23.02.0001.01)]{wolfram}.

\section{System Model}

\begin{figure}[h]
\begin{center}
\hspace*{-.4cm}\includegraphics[scale=0.24]{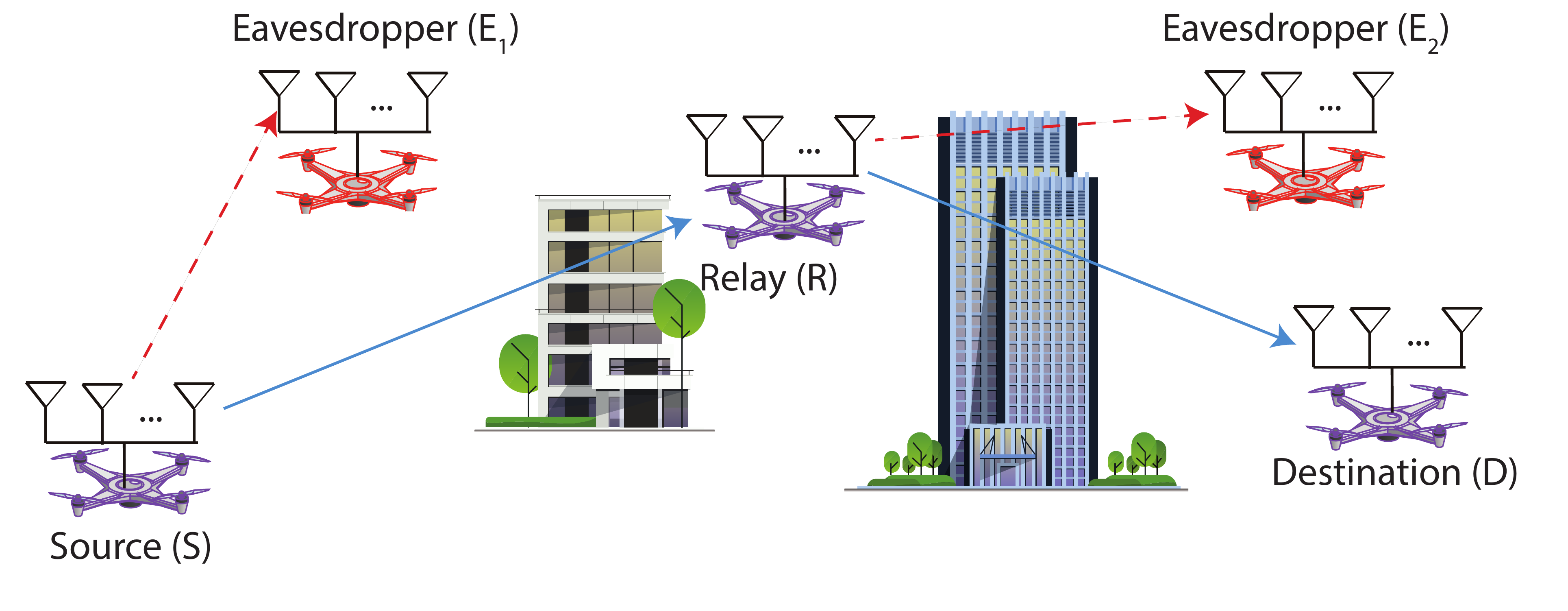}
\end{center}
\caption{System model.}
\end{figure}

We consider a dual-hop UAV-based WCS\ consisting of a source node $(S)$
connected to a destination $(D)$ via a relay $(R)$ employing DF\ protocol.
Two eavesdroppers, $E_{1}$ and $E_{2}$, try to intercept
the legitimate messages broadcasted by $S$ and $R$, in the first and second
time slots, respectively. We assume that the transmit nodes ($S$, $R$) are equipped
with a single transmit antenna while all receiving nodes ($R,$ $D$, $%
E_{1},$ and $E_{2}$) are equipped with $N_{R},$ $N_{D},$ $N_{E_{1}},$ and $%
N_{E_{2}}$ receiving antennas, respectively. Lastly, due to the limited UAV\
transmit power and the significant distance between $S$ and $D$, we assume
that there are no direct links from $S$ to $D$ and $E_{2}$. Without loss of
generality, the received signals at $R$, $D,$ $E_{1},$ and $E_{2}$ are
expressed as
\begin{equation}
y_{U}=\mathbf{h}_{SU}x+\mathbf{n}_{U},U\in \{R,E_{1}\},
\end{equation}%
\begin{equation}
y_{V}=\mathbf{h}_{RV}x^{\prime }+\mathbf{n}_{V},V\in \{D,E_{2}\},
\end{equation}%
where $x$ is the transmit signal, $x^{\prime }$ is the decoded and
regenerated one by $R$, $\mathbf{n}_{U}$ and $\mathbf{n}_{V}$ are the
additive white Gaussian noise (AWGN) realizations at either $R$, $D,$ $E_{1}$
or $E_{2}$'s antennas, which are independent and identically distributed
(i.i.d) zero-mean complex Gaussian random variables (ZMCGRVs) with
distribution $\mathcal{CN}\left( 0,\sigma _{n_{Z}}^{2}\right) $, with $Z\in
\left\{ U,V\right\}$. Additionally,
\begin{equation}
\mathbf{h}_{SU}=\left[ h_{SU}^{(1)},...,h_{SU}^{(N_{U})}\right] ^{T}
\end{equation}
and
\begin{equation}
\mathbf{h}_{RV}=\left[ h_{RV}^{(1)},...,h_{RV}^{(N_{V})}\right] ^{T}
\end{equation}%
\noindent are the $S$-$U$ and $R$-$V$ channel fading vector between $S/R$'s
transmit antenna and $U/V$'s receive ones, whose elements are complex-valued
i.i.d random variables with a fading envelope assumed to follow Nakagami-$m$
distribution with fading parameter $m_{XZ}$ and average fading power $\Omega
_{XZ}=\mathbb{E}\left[ \left\vert h_{XZ}^{(i)}\right\vert ^{2}\right] .$

Due to the UAV nodes' mobility, the actual channel fading coefficients ($%
\mathbf{h}_{SU}$ and $\mathbf{h}_{RV}$) are different from the estimated
one. Such a relationship can be expressed through the widely-known
first-order autoregressive process, $AR(1)$, as follows \cite%
{khattabi,anshul}%
\begin{equation}
\mathbf{h}_{XZ}=\rho _{XZ}\overline{\mathbf{h}}_{XZ}+\sqrt{1-\rho _{XZ}^{2}}%
\mathbf{w}_{XZ},
\end{equation}%
where $\rho _{XZ}$ is the underlying correlation coefficient, with $XZ\in
\left\{ SR,\text{ }SE_{1},\text{ }RD,\text{ }RE_{2}\right\} $, $\overline{%
\mathbf{h}}_{XZ}$ is the exact outdated CSI vector, and $\mathbf{w}_{XZ}$ is
the mobility\ noise vector whose entries are i.i.d ZMCGRVs with distribution
$\mathcal{CN}\left( 0,\sigma _{w_{XZ}}^{2}\right) $. On the other hand, the
correlation coefficient $\rho _{XZ}$ is defined, according to the Jakes'
model, as \cite{rosa,khattabi}%
\begin{equation}
\rho _{XZ}=J_{0}\left( \frac{2\pi v_{XZ}f_{c}\tau }{c}\right) ,  \label{corr}
\end{equation}%
which is a function of the two nodes' relative speed $v_{XZ}$ along with the
delay time $\tau $ between the CSI\ estimation and signal reception, and the
carrier frequency $f_{c}.\ $Moreover, in addition to the channel\
decorrelation over time, its estimation is also subject to the inherent
noise in the receiver's tracking loop. The final expression of
the channel fading coefficient can be expressed as \cite{khattabi}%
\begin{equation}
\mathbf{h}_{XZ}=\rho _{XZ}\underset{\overline{\mathbf{h}}_{XZ}}{\underbrace{%
\left( \mathbf{\hat{h}}_{XZ}+\mathbf{\varepsilon }_{XZ}\right) }}+\sqrt{%
1-\rho _{XZ}^{2}}\mathbf{w}_{XZ},
\end{equation}%
where $\mathbf{\hat{h}}_{XZ}$ is the estimated channel vector, and $\mathbf{%
\varepsilon }_{XZ}$ is the estimation noise vector with i.i.d ZMCGRVs
entries with distribution $\mathcal{CN}\left( 0,\sigma _{\varepsilon
_{XZ}}^{2}\right) $.

\begin{remark}
The channel's mobility-dependent correlation coefficient $\left( \text{i.e.,
}\rho _{XZ}\right) ,$ given in (\ref{corr}), manifests the impact of nodes'
mobility on the time-varying statistical behavior of the channel. Such
coefficient is expressed in terms of the nodes' relative speed, carrier
frequency, delay, and speed of light. A static scenario, i.e., $%
v_{XZ}=0$, corresponds to $\rho _{XZ}=1,$ which refers to a perfect
correlation case. Likewise, at very high $v_{XZ}$ and/or $f_{c}$ and/or $%
\tau $, $\rho _{XZ}$ tends to 0, yielding severe channel time variations, in
which the estimated CSI\ is completely uncorrelated with the actual one.
Also, it worth mentioning that $J_{0}\left( x\right) $ in (\ref{corr})$,$
with $x=\frac{2\pi v_{XZ}f_{c}\tau }{c},$ is a decreasing function of $%
v_{XZ} $ and/or $f_{c}$ and/or $\tau $ over $x\in \left[ 0,x_{0}\right] ,$
with $x_{0}=3.8317$ being the function's minima that can be found
numerically by retrieving the first zero of $\left( J_{0}\left( x\right)
\right) ^{\prime }$.
\end{remark}

At the legitimate and wiretap nodes (i.e., $R$, $D,$ $E_{1}$, or $E_{2})$,
the MRC\ receiver is used to combine the received signals copies on its
various diversity branches. To this end, the combined signal at the output
of the MRC\ receiver at such nodes can be formulated as%
\begin{align}
y_{Z}^{(MRC)}& =\rho _{XZ}\mathbf{\phi }_{XZ}\mathbf{\hat{h}}_{XZ}u+\rho
_{XZ}\mathbf{\phi }_{XZ}\mathbf{\varepsilon }_{XZ}u  \notag \\
& +\mathbf{\phi }_{XZ}\sqrt{1-\rho _{XZ}^{2}}\mathbf{w}_{XZ}u+\mathbf{\phi }%
_{XZ}\mathbf{n}_{Z},
\end{align}%
for $XZ\in \left\{ SR,\text{ }RD,\text{ }SE_{1},\text{ }RE_{2}\right\} $,
where
\begin{equation}
u=\left\{
\begin{array}{l}
x,\text{ if }Z\in \left\{ R,E_{1}\right\} \\
x^{\prime },\text{ if }Z\in \left\{ D,E_{2}\right\}%
\end{array}%
\right. ,
\end{equation}%
and $\mathbf{\phi }_{XZ}=\frac{\left( \mathbf{\hat{h}}_{XZ}\right) ^{H}}{%
\left\Vert \mathbf{\hat{h}}_{XZ}\right\Vert _{2}}$ denotes the MRC$\ $%
beamforming vector. As a result, the corresponding instantaneous\ SNRs\ at
the output of the receiving nodes' combiners can be written using the
following generalized form

\begin{eqnarray}
\gamma _{XZ} &=&\frac{\rho _{XZ}^{2}P_{X}\left\Vert \mathbf{\hat{h}}%
_{XZ}\right\Vert _{2}^{2}}{P_{X}\left( \rho _{XZ}^{2}\sigma _{\varepsilon
_{XZ}}^{2}+\left( 1-\rho _{XZ}^{2}\right) \sigma _{w_{XZ}}^{2}\right)
+\sigma _{n_{Z}}^{2}},  \notag \\
&=&\sum\limits_{i=1}^{N_{Z}}\gamma _{XZ}^{(i)},  \label{snrtotal}
\end{eqnarray}%
where $P_{X}$ is the transmit power of node $X$, and
\begin{equation}
\gamma _{XZ}^{(i)}=\frac{P_{X}\rho _{XZ}^{2}\left\vert \hat{h}%
_{XZ}^{(i)}\right\vert ^{2}}{P_{X}\left( \rho _{XZ}^{2}\sigma _{\varepsilon
_{XZ}}^{2}+\left( 1-\rho _{XZ}^{2}\right) \sigma _{w_{XZ}}^{2}\right)
+\sigma _{n_{Z}}^{2}}
\end{equation}%
denotes the instantaneous SNR\ at the $i$th branch, with an average value
given as%
\begin{equation}
\Upsilon _{XZ}=\frac{\rho _{XZ}^{2}\delta _{XZ}}{\frac{\delta _{XZ}}{\Omega
_{XZ}}\left( \rho _{XZ}^{2}\sigma _{\varepsilon _{XZ}}^{2}+\left( 1-\rho
_{XZ}^{2}\right) \sigma _{w_{XZ}}^{2}\right) +1},  \label{avgsnrbranch}
\end{equation}%
with $\delta _{XZ}=\frac{P_{X}\Omega _{XZ}}{\sigma _{n_{Z}}^{2}}.$

\begin{remark}

\begin{enumerate}
\item By differentiating the average SNR\ per branch in (\ref{avgsnrbranch})
with respect to $\rho _{XZ}$, it yields%
\begin{equation}
\frac{\partial \Upsilon _{XZ}}{\partial \rho _{XZ}}=\frac{2\rho _{XZ}\delta
_{XZ}\left[ 1+\frac{\delta _{XZ}\sigma _{w_{XZ}}^{2}}{\Omega _{XZ}}\right] }{%
\left( \frac{\delta _{XZ}}{\Omega _{XZ}}\left( \rho _{XZ}^{2}\sigma
_{\varepsilon _{XZ}}^{2}+\left( 1-\rho _{XZ}^{2}\right) \sigma
_{w_{XZ}}^{2}\right) +1\right) ^{2}},
\end{equation}%
which is strictly positive. Henceforth, we can conclude that the effective
per-branch average SNR\ is an increasing function of the correlation
coefficient $\rho _{XZ}.$
\item On the other hand, one can note from (\ref{avgsnrbranch}) that at
higher average SNR values (i.e., $\delta _{XZ}\rightarrow \infty $), the
effective average SNR per branch\ $\left( \text{i.e., }\Upsilon _{XZ}\right)
$ converges to the following ceiling value
\begin{equation}
\Upsilon _{XZ}^{\left( \infty \right) }=\frac{\rho _{XZ}^{2}\Omega _{XZ}}{%
\left( \rho _{XZ}^{2}\sigma _{\varepsilon _{XZ}}^{2}+\left( 1-\rho
_{XZ}^{2}\right) \sigma _{w_{XZ}}^{2}\right) }.
\end{equation}
The above upper bound of $\Upsilon _{XZ}$ leads to saturation floors in the
system's IP at high $\delta _{XZ}$ values, within which the IP will be
depending essentially on $\rho _{XZ}^{2},$ $\sigma _{\varepsilon _{XZ}}^{2},$
and $\sigma _{w_{XZ}}^{2}$ values. As a consequence, we can conclude that
the impact of nodes mobility jointly with estimation imperfections can lead
to remarkable secrecy performance limitations which need to be considered in
the design of such systems.
\end{enumerate}
\end{remark}

\section{Statistical Properties}

In this section, statistical properties such as the probability density
function (PDF) and the cumulative distribution function (CDF) of the per-hop
SNR\ are presented.

From (\ref{snrtotal}), one can infer that the instantaneous SNRs are
proportional to the sum of Gamma-distributed squared fading envelopes at
each branch of the receiver. To this end, the respective PDF\ and CDF\ of
such instantaneous SNRs can be written as \cite{access}%
\begin{equation}
f_{\gamma _{XZ}}\left( y\right) =\frac{\left( \frac{m_{XZ}}{\Upsilon _{XZ}}%
\right) ^{m_{XZ}N_{Z}}}{\Gamma \left( m_{XZ}N_{Z}\right) }%
y^{m_{XZ}N_{Z}-1}\exp \left( -\frac{m_{XZ}}{\Upsilon _{XZ}}y\right) ,
\label{pdf}
\end{equation}%
and%
\begin{equation}
F_{\gamma _{SE_{1}}}\left( y\right) =\frac{\gamma _{\text{inc}}\left(
m_{XZ}N_{Z},\frac{m_{XZ}}{\Upsilon _{XZ}}y\right) }{\Gamma \left(
m_{XZ}N_{Z}\right) },  \label{cdf}
\end{equation}%
respectively.

\section{Secrecy Analysis}

In this section, exact closed-form and asymptotic expressions of the IP\ are
derived, in terms of the main setup parameters.

The IP metric is defined as the probability that the secrecy capacity,
which is the difference between the capacity of the legitimate channels
and that of the eavesdropping ones, is less than or equal to zero.
Mathematically, it is defined as \cite{ojcoms} {{\
\begin{align}
P_{int}& =\Pr \left( C_{s}\leq 0\right)  \notag \\
& =1-\Pr \left( C_{s}>0\right) ,  \label{ipnew}
\end{align}%
where}}

\begin{equation}
C_{s}=\min \left( C_{s}^{(SR)},C_{s,eq}^{(RD)}\right) ,  \label{cs1}
\end{equation}%
\begin{equation}
C_{s}^{(SR)}=\log _{2}\left( \frac{1+\gamma _{SR}}{1+\gamma _{SE_{1}}}%
\right) ,  \label{cs11}
\end{equation}%
and%
\begin{equation}
C_{S,eq}^{(RD)}=\min \left( C_{s}^{(SRD)},C_{s}^{(RD)}\right) ,  \label{cseq}
\end{equation}%
refer to the end-to-end, first hop, and second hops' secrecy capacities,
respectively, by assuming DF\ relaying protocol, with%
\begin{eqnarray}
C_{s}^{(SRD)} &=&\log _{2}\left( \frac{1+\gamma _{SR}}{1+\gamma _{RE_{2}}}%
\right) ,  \label{cs12} \\
C_{s}^{(RD)} &=&\log _{2}\left( \frac{1+\gamma _{RD}}{1+\gamma _{RE_{2}}}%
\right) .  \label{cs2}
\end{eqnarray}

{To this end, the total secrecy capacity in (\ref{cs1}) can be formulated as
follows}%
\begin{equation}
C_{s}=\min \left( C_{s}^{(SR)},C_{s}^{(RD)},C_{s}^{(SRD)}\right) .
\label{csmin}
\end{equation}

\noindent Leveraging the probability theory, (\ref{ipnew}) becomes
\begin{align}
P_{int}& =1-\Pr \left( C_{s}>0,\gamma _{SR}>\gamma _{th}\right)  \notag \\
& -\Pr \left( C_{s}>0,\gamma _{SR}<\gamma _{th}\right) ,  \label{ipinter}
\end{align}%
where $\gamma _{th}$ is a decoding threshold SNR, below which the decoding
process can not be performed at the relay.

When $\gamma _{SR}<\gamma _{th},$ the relay fails at decoding the information
message. Therefore, no signal will be transmitted to $D$ as well as the
eavesdropper $E_{2}.$\ Hence, we have $\gamma _{RD}=\gamma _{RE_{2}}=0$,
which yields from (\ref{cseq}), (\ref{cs12}), and (\ref{cs2}) that $%
C_{s,eq}^{(RD)}\leq 0,$ and consequently,{\ it yields from (\ref{cs1}) that }%
$C_{s}\leq 0$ and $\Pr \left( C_{s}>0,\gamma _{SR}<\gamma _{th}\right) =0.$
As a result, the overall\ IP\ expression in {(\ref{ipinter})} reduces to%
\begin{equation}
P_{int}=1-\Pr \left( C_{s}>0,\gamma _{SR}>\gamma _{th}\right) .
\label{pinttt}
\end{equation}

\begin{remark}
\begin{enumerate}
\item {{From{\ {{(\ref{csmin}),}} }}}it can be noticed that the system's
secrecy capacity is the minimum of three elementary secrecy capacities,
namely the first hop's one; i.e., $C_{s}^{(SR)}$; the second's; i.e., $%
C_{s}^{(RD)}$; and a cross-term one; i.e., $C_{s}^{(SRD)}$; which involves
the first hop's legitimate SNR\ along with the second hop's wiretap one.
Indeed, the higher are these three terms, the greater the overall system's
secrecy capacity. Notably, {{{one can note from (\ref{cs11}), (\ref{cs12}),
and (\ref{cs2}) that such elementary secrecy capacities, and consequently
the overall one in {{{{(\ref{csmin})}}}}, increase with the rise of the
legitimate instantaneous per-hop SNRs, i.e., }}}$\gamma _{SR}$ and $\gamma
_{RD}$ (better system secrecy), and decrease versus the wiretap ones, {i.e.,
}$\gamma _{SE_{1}}$ and $\gamma _{RE_{2}}$ (worse system secrecy). Thus, the
IP, given by {{{(\ref{ipnew}) is a decreasing function of }}}$\gamma _{SR}$
and $\gamma _{RD}${, and increasing with respect to }$\gamma _{SE_{1}}$ and $%
\gamma _{RE_{2}}.$

\item The secrecy capacities in {{{(\ref{cs11}), (\ref{cs12}), and (\ref{cs2}%
)}}} can be written in a generalized form as%
\begin{equation}
C_{s}^{\left( \xi \right) }=\log _{2}\left( \frac{1+\varrho _{l}g_{l}}{%
1+\varrho _{w}g_{w}}\right) ,  \label{cshop}
\end{equation}%
with $\varrho _{x}=\frac{\Upsilon _{x}}{\Omega _{x}},x\in \left\{
l,w\right\} ,$
\begin{equation}
\left( \xi ,l,w\right) \in \left\{
\begin{array}{l}
\left( SR,SR,SE_{1}\right) , \\
\left( SRD,SR,RE_{2}\right) , \\
\left( RD,RD,RE_{2}\right)%
\end{array}%
\right\} ,
\end{equation}
and%
\begin{equation}
g_{x}=\sum\limits_{i=1}^{N^{(x)}}\left\vert h_{x}^{(i)}\right\vert ^{2},
\label{gx}
\end{equation}%
with $N^{(x)}=\left\{
\begin{array}{l}
N_{R},\left( x=l=SR\right) \\
N_{D},\left( x=l=RD\right) \\
N_{E_{1}},\left( x=w=SE_{1}\right) \\
N_{E_{2}},\left( x=w=RE_{2}\right)%
\end{array}%
\right. .$

Thus, as mentioned in \textit{Remark 3.1}, a secure system from the PLS\
perspective corresponds to a higher secrecy capacity. To this end, it can be
intuitively noticed from {{{(\ref{cshop}) and}}} {{{(\ref{gx})}}} that the
higher $\varrho _{l}$ and/or the number of legitimate nodes' antennas (i.e.,
$N_{R},N_{D}$), the greater is $C_{s}^{\left( \xi \right) }$. Likewise, this
latter drops by increasing $\varrho _{w}$ and/or $N_{E_{1}},$ $N_{E_{2}}$.
Therefore, to gain more insights, we inspect the behavior of $C_{s}^{\left(
\xi \right) }$, for equal average SNRs, i.e., $\varrho _{l}=\varrho
_{w}=\varrho $, and number of antennas, i.e., $N^{(l)}=N^{(w)}$, by
differentiating $C_{s}^{\left( \xi \right) }$ with respect to $\varrho $ as%
\begin{equation}
\frac{\partial C_{s}^{\left( \xi \right) }}{\partial \varrho }=\frac{%
g_{l}-g_{w}}{\left( 1+\varrho g_{l}\right) \left( 1+\varrho g_{w}\right) }.
\label{deriv}
\end{equation}%
One can note from {(\ref{deriv})} that $C_{s}^{\left( \xi \right) }$ is an
increasing function of $\varrho $ iff $g_{l}-g_{w}>0.$ Notably, as this
last-mentioned quantity is random, we are interested then into analyzing the
probability$:\ \Pr \left[ g_{l}-g_{w}>0\right] .$\ Hence, it yields
\begin{eqnarray}
\mathcal{Y} &=&\Pr \left[ g_{l}-g_{w}>0\right] ,  \notag \\
&=&\int_{0}^{\infty }F_{g_{w}}\left( y\right) f_{g_{l}}\left( y\right) dy,
\end{eqnarray}%
where $F_{g_{w}}\left( .\right) $ and $f_{g_{l}}\left( .\right) $ are given,
respectively, as follows%
\begin{equation}
F_{g_{w}}\left( y\right) =\frac{\gamma _{\text{inc}}\left( m_{w}N^{\left(
w\right) },\frac{m_{w}}{\Omega _{w}}y\right) }{\Gamma \left( m_{w}N^{\left(
w\right) }\right) },  \label{cdfg1}
\end{equation}%
\begin{equation}
f_{g_{l}}\left( y\right) =\frac{\left( \frac{m_{l}}{\Omega _{l}}\right)
^{m_{l}N^{(l)}}}{\Gamma \left( m_{l}N^{\left( l\right) }\right) }%
y^{m_{l}N^{\left( l\right) }-1}\exp \left( -\frac{m_{l}}{\Omega _{l}}%
y\right) .  \label{pdfg1}
\end{equation}%
Thus, it produces
\begin{align}
\mathcal{Y}& \mathcal{=}\frac{\left( \frac{m_{l}}{\Omega _{l}}\right)
^{m_{l}N^{\left( l\right) }}}{\Gamma \left( m_{l}N^{\left( l\right) }\right)
\Gamma \left( m_{w}N^{\left( w\right) }\right) }\int_{0}^{\infty }\gamma _{%
\text{inc}}\left( m_{w}N^{(w)},\frac{m_{w}}{\Omega _{w}}y\right)  \notag \\
& \times y^{m_{l}N^{\left( l\right) }-1}\exp \left( -\frac{m_{l}}{\Omega _{l}%
}y\right) dy.
\end{align}%
Through the use of \cite[Eq. (6.455.1)]{integrals} , one obtains {(\ref{Y})
given at the top of the next page.}

Figs. \ref{fig1a} and \ref{fig1b} present the plot of {(\ref{Y}) versus }$%
m_{l}$ and $m_{w}$ (Fig. \ref{fig1a}), and $\Omega _{l}$ and $\Omega _{w}$
(Fig. \ref{fig1b}), with $N^{(l)}=N^{(w)}=3$. It can be noted from Fig. \ref%
{fig1a} that $\mathcal{Y}$ slightly increases versus $m_{l}$ and decreases
versus $m_{w}.$ Essentially, for equal number of antennas, this probability
does not exceed $56\%$ for $m_{x}\leq 10$. Nevertheless, as seen from Fig. %
\ref{fig1b}, $\mathcal{Y}$ (i.e., the probability that $C_{s}^{\left( \xi
\right) }$ is increasing vs $\varrho $) surpasses $99$\% for $\Omega
_{l}\geq 4\Omega _{w}.$%
\begin{figure*}[t]
{\normalsize 
\setcounter{mytempeqncnt}{\value{equation}}
\setcounter{equation}{33} }
\par
\begin{equation}
\mathcal{Y}\mathcal{=}\frac{\left( \frac{m_{l}}{\Omega _{l}}\right)
^{m_{l}N^{\left( l\right) }}\left( \frac{m_{w}}{\Omega _{w}}\right)
^{m_{w}N^{\left( w\right) }}\Gamma \left( m_{l}N^{\left( l\right)
}+m_{w}N^{\left( w\right) }\right) }{\Gamma \left( m_{l}N^{\left( l\right)
}\right) \Gamma \left( m_{w}N^{\left( w\right) }+1\right) \left( \frac{m_{w}%
}{\Omega _{w}}+\frac{m_{l}}{\Omega _{l}}\right) ^{m_{l}N^{\left( l\right)
}+m_{w}N^{\left( w\right) }}}\text{ }_{2}F_{1}\left( 1,m_{l}N^{\left(
l\right) }+m_{w}N^{\left( w\right) };m_{w}N^{\left( w\right) }+1;\frac{\frac{%
m_{w}}{\Omega _{w}}}{\frac{m_{w}}{\Omega _{w}}+\frac{m_{l}}{\Omega _{l}}}%
\right) .  \label{Y}
\end{equation}%
\par
{\normalsize 
\hrulefill 
\vspace*{4pt} }
\end{figure*}
\begin{figure}[tbp]
\begin{center}
\hspace*{-1cm}%
\includegraphics[scale=0.64]{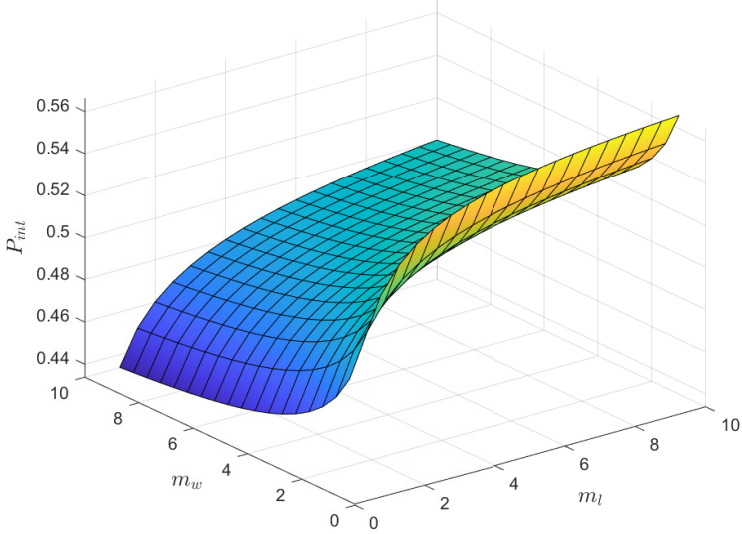}
\end{center}
\caption{IP versus $m_{l}$ and $m_{w}$, for $\Omega _{l}=\Omega _{w}=2.5$, $%
N^{(l)}=N^{(w)}=3$.}
\label{fig1a}
\end{figure}
\begin{figure}[h]
\begin{center}
\hspace*{-1cm} %
\includegraphics[scale=0.64]{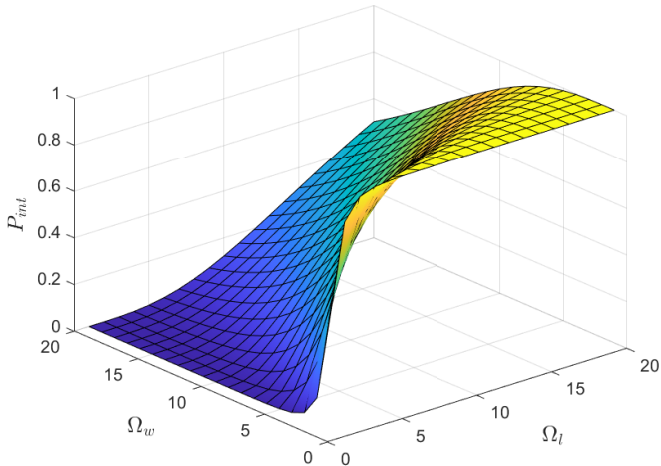}
\end{center}
\caption{IP versus $\Omega _{l}$ and $\Omega _{w}$, for $m_{l}=m_{w}=2$, $%
N^{(l)}=N^{(w)}=3$.}
\label{fig1b}
\end{figure}
\end{enumerate}
\end{remark}

\subsection{Exact Analysis}

\begin{proposition}
The IP\ of the considered dual-hop UAV-based WCS\ is given in closed-form
expression as%
\begin{align}
P_{int}& =1-\frac{\left( \frac{m_{SR}}{\Upsilon _{SR}}\right)
^{m_{SR}N_{R}}\left( \frac{m_{RE_{2}}}{\Upsilon _{RE_{2}}}\right)
^{m_{RE_{2}}N_{E_{2}}}}{\left( m_{SR}N_{R}-1\right) !\left(
m_{RE_{2}}N_{E_{2}}-1\right) !}  \notag \\
& \times \sum_{l=0}^{m_{RD}N_{D}-1}\frac{\left(
m_{RE_{2}}N_{E_{2}}+l-1\right) !\left( \frac{m_{RD}}{\Upsilon _{RD}}\right)
^{l}}{l!\left[ \frac{m_{RE_{2}}}{\Upsilon _{RE_{2}}}+\frac{m_{RD}}{\Upsilon
_{RD}}\right] ^{m_{RE_{2}}N_{E_{2}}+l}}\sum_{i=1}^{4}\mathcal{I}^{(i)},
\label{ipfinal}
\end{align}%
where%
\begin{equation}
\mathcal{I}^{(1)}=\frac{\Gamma _{\text{inc}}\left( m_{SR}N_{R},\frac{m_{SR}}{%
\Upsilon _{SR}}\gamma _{th}\right) }{\left( \frac{m_{SR}}{\Upsilon _{SR}}%
\right) ^{m_{SR}N_{R}}},  \label{I1}
\end{equation}%
\begin{align}
\mathcal{I}^{(2)}& =-\sum_{n=0}^{m_{SE_{1}}N_{E_{1}}-1}\frac{\left( \frac{%
m_{SE_{1}}}{\Upsilon _{SE_{1}}}\right) ^{n}}{\left[ \frac{m_{SR}}{\Upsilon
_{SR}}+\frac{m_{SE_{1}}}{\Upsilon _{SE_{1}}}\right] ^{m_{SR}N_{R}+n}n!}
\notag \\
& \times \Gamma _{\text{inc}}\left( m_{SR}N_{R}+n,\left[ \frac{m_{SR}}{%
\Upsilon _{SR}}+\frac{m_{SE_{1}}}{\Upsilon _{SE_{1}}}\right] \gamma
_{th}\right) ,  \label{I2}
\end{align}%
\begin{align}
\mathcal{I}^{(3)}& =-\sum_{p=0}^{m_{RE_{2}}N_{E_{2}}+l-1}\frac{\left( \frac{%
m_{RE_{2}}}{\Upsilon _{RE_{2}}}+\frac{m_{RD}}{\Upsilon _{RD}}\right) ^{p}}{p!%
}  \notag \\
& \times \frac{\Gamma _{\text{inc}}\left( m_{SR}N_{R}+p,\Delta \gamma
_{th}\right) }{\Delta ^{m_{SR}N_{R}+p}},  \label{I3}
\end{align}%
\begin{align}
\mathcal{I}^{(4)}& =\sum_{n=0}^{m_{SE_{1}}N_{E_{1}}-1}\left(
\sum_{p=0}^{m_{RE_{2}}N_{E_{2}}+l-1}\frac{\left( \frac{m_{RE_{2}}}{\Upsilon
_{RE_{2}}}+\frac{m_{RD}}{\Upsilon _{RD}}\right) ^{p}}{n!p!}\right.  \notag \\
& \times \left. \frac{\Gamma _{\text{inc}}\left( m_{SR}N_{R}+n+p,\Psi \gamma
_{th}\right) \left( \frac{m_{SE_{1}}}{\Upsilon _{SE_{1}}}\right) ^{n}}{\Psi
^{m_{SR}N_{R}+n+p}}\right) ,  \label{I4}
\end{align}%
$\Delta =\frac{m_{RE_{2}}}{\Upsilon _{RE_{2}}}+\frac{m_{RD}}{\Upsilon _{RD}}+%
\frac{m_{SR}}{\Upsilon _{SR}},$ and $\Psi =\Delta +\frac{m_{SE_{1}}}{%
\Upsilon _{SE_{1}}}.$

\begin{proof}
The proof is provided in Appendix A.
\end{proof}
\end{proposition}

\begin{remark}
\begin{enumerate}
\item The IP\ closed-form expression, given by (\ref{ipfinal})-(\ref{I4}) is
formulated in terms of finite summations, upper-incomplete and complete
Gamma functions, and exponential terms. All the system and channel
parameters are involved in such functions, such as Nakagami-$m$ fading
parameter $\left( m_{XZ}\right) $, the threshold SNR\ $\gamma _{th}$, and
effective average SNRs $\left( \Upsilon _{XZ}\right) .$ These latter are
essentially expressed, as shown in (\ref{avgsnrbranch}), in terms of the
respective mobility-dependent correlation coefficients $\left( \rho
_{XZ}\right) $, computed based on UAVs' relative speed, carrier frequency,
and delay, average fading power $\left( \Omega _{XZ}\right) $, and noise
powers due to the CSI estimation errors and mobility $\left( \sigma
_{\varepsilon _{XZ}}^{2},\sigma _{w_{XZ}}^{2}\right) $. Of note, the
aforementioned functions are already implemented in almost all computational
software (e.g., MATLAB, MATHEMATICA), which can provide insightful
observations on the impact of the system parameters on the overall secrecy
performance.

\item One can obviously note that at lower threshold SNR values; i.e., $%
\gamma _{th}\rightarrow 0$; the terms (\ref{I1})-(\ref{I4}), leveraging $%
\Gamma _{\text{inc}}\left( a,0\right) =\Gamma \left( a\right) ,$ reduce to
\begin{equation}
\mathcal{V}^{(1)}=\frac{\left( m_{SR}N_{R}-1\right) !}{\left( \frac{m_{SR}}{%
\Upsilon _{SR}}\right) ^{m_{SR}N_{R}}},  \label{v1}
\end{equation}%
\begin{equation}
\mathcal{V}^{(2)}=-\sum_{n=0}^{m_{SE_{1}}N_{E_{1}}-1}\frac{\left( \frac{%
m_{SE_{1}}}{\Upsilon _{SE_{1}}}\right) ^{n}\left( m_{SR}N_{R}+n-1\right) !}{%
n!\left[ \frac{m_{SR}}{\Upsilon _{SR}}+\frac{m_{SE_{1}}}{\Upsilon _{SE_{1}}}%
\right] ^{m_{SR}N_{R}+n}},  \label{v2}
\end{equation}%
\begin{align}
\mathcal{V}_{l}^{(3)}& =-\sum_{p=0}^{m_{RE_{2}}N_{E_{2}}+l-1}\frac{\left(
m_{SR}N_{R}+p-1\right) !}{p!}  \notag \\
& \times \frac{\left( \frac{m_{RE_{2}}}{\Upsilon _{RE_{2}}}+\frac{m_{RD}}{%
\Upsilon _{RD}}\right) ^{p}}{\Delta ^{m_{SR}N_{R}+p}},  \label{v3}
\end{align}%
and%
\begin{align}
\mathcal{V}_{l}^{(4)}&
=\sum_{n=0}^{m_{SE_{1}}N_{E_{1}}-1}\sum_{p=0}^{m_{RE_{2}}N_{E_{2}}+l-1}\frac{%
\left( m_{SR}N_{R}+n+p-1\right) !}{\Psi ^{m_{SR}N_{R}+n+p}}  \notag \\
& \times \frac{\left( \frac{m_{SE_{1}}}{\Upsilon _{SE_{1}}}\right)
^{n}\left( \frac{m_{RE_{2}}}{\Upsilon _{RE_{2}}}+\frac{m_{RD}}{\Upsilon _{RD}%
}\right) ^{p}}{n!p!},  \label{v4}
\end{align}%
which are independent of $\gamma _{th}$. Henceforth, in this regime, the
system's IP\ is not improved regardless of the decrease in $\gamma _{th}$
(i.e., increasing the decoding probability at the relay), as the IP\ depends
exclusively on the per-node number of antennas, links' fading parameters,
and average SNRs.

\item On the other hand, by making use of \cite[Eq. (8.350.4)]{integrals},
i.e., $\Gamma _{\text{inc}}\left( a,\infty \right) =0$ $\forall a>0$, the
terms (\ref{I1})-(\ref{I4}) vanish for high $\gamma _{th}$ values (i.e., $%
\gamma _{th}\rightarrow \infty $). Henceforth, the IP, given by (\ref%
{ipfinal}), reaches one. In fact, the greater $\gamma _{th}$, the lower is
the probability of successful decoding at the relay. As such, the relay
fails at relaying the information signal to $D$, which results in zero
capacity/SNR of the second hop.
\end{enumerate}
\end{remark}

\subsection{Asymptotic Analysis}

In this subsection, asymptotic expressions for the system's IP\ are derived
at high legitimate links' average SNR\ values. Two scenarios
are investigated, namely (i)\ moving nodes with
imperfect channel estimation, and (ii) static nodes with
perfect CSI at the receiver.

\subsubsection{Scenario I\ (Moving UAVs with Imperfect CSI)}

In this scenario, when the average SNRs $\delta _{SR}$ and $\delta _{RD}$
become high (i.e., $\delta _{SR},$ $\delta _{RD}\rightarrow \infty $), their
corresponding effective ones, given by (\ref{avgsnrbranch}), reduce to%
\begin{equation}
\Xi _{XZ}\sim \frac{\rho _{XZ}^{2}\Omega _{XZ}}{\rho _{XZ}^{2}\sigma
_{\varepsilon _{XZ}}^{2}+\left( 1-\rho _{XZ}^{2}\right) \sigma _{w_{XZ}}^{2}}%
,XZ\in \left\{ SR,RD\right\} .
\end{equation}

As a consequence, by substituting $\Upsilon _{XZ}$ $\left( XZ\in \left\{
SR,RD\right\} \right) $ in (\ref{ipfinal})-(\ref{I4}) by $\Xi _{XZ}$, given
by the above equation, the system's IP\ in such a particular case can be
expressed as
\begin{align}
P_{int}^{\left( \infty ,1\right) }& \sim 1-\frac{\left( \frac{m_{SR}}{\Xi
_{SR}}\right) ^{m_{SR}N_{R}}\left( \frac{m_{RE_{2}}}{\Upsilon _{RE_{2}}}%
\right) ^{m_{RE_{2}}N_{E_{2}}}}{\left( m_{SR}N_{R}-1\right) !\left(
m_{RE_{2}}N_{E_{2}}-1\right) !}  \notag \\
\times & \sum_{l=0}^{m_{RD}N_{D}-1}\frac{\left(
m_{RE_{2}}N_{E_{2}}+l-1\right) !\left( \frac{m_{RD}}{\Xi _{RD}}\right) ^{l}}{%
l!\left[ \frac{m_{RE_{2}}}{\Upsilon _{RE_{2}}}+\frac{m_{RD}}{\Xi _{RD}}%
\right] ^{m_{RE_{2}}N_{E_{2}}+l}}\sum_{i=1}^{4}\mathcal{H}^{(i)},
\label{ipass1}
\end{align}%
where%
\begin{equation}
\mathcal{H}^{(1)}=\frac{\Gamma _{\text{inc}}\left( m_{SR}N_{R},\frac{m_{SR}}{%
\Xi _{SR}}\gamma _{th}\right) }{\left( \frac{m_{SR}}{\Xi _{SR}}\right)
^{m_{SR}N_{R}}},  \label{H1}
\end{equation}%
\begin{align}
\mathcal{H}^{(2)}& =-\sum_{n=0}^{m_{SE_{1}}N_{E_{1}}-1}\frac{\left( \frac{%
m_{SE_{1}}}{\Upsilon _{SE_{1}}}\right) ^{n}}{n!}  \notag \\
& \times \frac{\Gamma _{\text{inc}}\left( m_{SR}N_{R}+n,\left[ \frac{m_{SR}}{%
\Xi _{SR}}+\frac{m_{SE_{1}}}{\Upsilon _{SE_{1}}}\right] \gamma _{th}\right)
}{\left[ \frac{m_{SR}}{\Xi _{SR}}+\frac{m_{SE_{1}}}{\Upsilon _{SE_{1}}}%
\right] ^{m_{SR}N_{R}+n}},  \label{H2}
\end{align}%
\begin{align}
\mathcal{H}^{(3)}& =-\sum_{p=0}^{m_{RE_{2}}N_{E_{2}}+l-1}\frac{\left( \frac{%
m_{RE_{2}}}{\Upsilon _{RE_{2}}}+\frac{m_{RD}}{\Xi _{RD}}\right) ^{p}}{p!}
\notag \\
& \times \frac{\Gamma _{\text{inc}}\left( m_{SR}N_{R}+p,\mathcal{Q}\gamma
_{th}\right) }{\mathcal{Q}^{m_{SR}N_{R}+p}},  \label{H3}
\end{align}%
\begin{align}
\mathcal{H}^{(4)}& =\sum_{p=0}^{m_{RE_{2}}N_{E_{2}}+l-1}\left(
\sum_{n=0}^{m_{SE_{1}}N_{E_{1}}-1}\frac{\left( \frac{m_{RE_{2}}}{\Upsilon
_{RE_{2}}}+\frac{m_{RD}}{\Xi _{RD}}\right) ^{p}}{p!}\right.  \notag \\
& \times \left. \frac{\left( \frac{m_{SE_{1}}}{\Upsilon _{SE_{1}}}\right)
^{n}\Gamma _{\text{inc}}\left( m_{SR}N_{R}+p+n,\mathcal{S}\gamma
_{th}\right) }{n!\mathcal{S}^{m_{SR}N_{R}+p+n}}\right) ,  \label{H4}
\end{align}%
with $\mathcal{Q}=\frac{m_{RE_{2}}}{\Upsilon _{RE_{2}}}+\frac{m_{RD}}{\Xi
_{RD}}+\frac{m_{SR}}{\Xi _{SR}},$ and $\mathcal{S}=\mathcal{Q}+\frac{%
m_{SE_{1}}}{\Upsilon _{SE_{1}}}.$

\begin{remark}
The IP's asymptotic expansion for Scenario I, given by (\ref{ipass1})-(%
\ref{H4}), is independent from $\delta _{SR}$ and $\delta _{RD}$.
For fixed values of the remainder of system and channel
parameters, the IP\ exhibits asymptotic floors at the high SNR regime.
Hence, the system's secrecy is not improved further regardless of the
increase in $\delta _{SR}$ and/or $\delta _{RD}.$ Consequently, the secrecy
diversity order in this scenario is equal to zero. Therefore, the system's
IP\ in the high SNR\ regime, for this scenario, depends exclusively on the
remainder of system parameters, namely the number of antennas $\left( N_{Z},%
\text{ }Z\in \{R,D,E_{1},E_{2}\}\right) ,$ fading parameters $\left( m_{XZ},%
\text{ }XZ\in \{SR,RD,SE_{1},RE_{2}\}\right) $, wiretap links' effective
average SNRs $\left( \Upsilon _{SE_{1}},\text{ }\Upsilon _{RE_{2}}\right) ,$
per-hop mobility-dependent correlation coefficients $\rho _{XZ},$ and the
noise powers due to channel estimation errors and mobility $\left( \sigma
_{\varepsilon _{XZ}}^{2},\text{ }\sigma _{w_{XZ}}^{2}\right) .$
\end{remark}

\subsubsection{Scenario II (Static Nodes with Perfect CSI)}

In this scenario, leveraging \textit{Remark 1}, the mobility-dependent
correlation coefficients, given in (\ref{corr}), equals 1. Furthermore,
under perfect channel estimation, we have $\sigma _{\varepsilon
_{XZ}}^{2}=0. $ To this end, the per-hop effective average SNR in (\ref%
{avgsnrbranch}) reduces to\ $\delta _{XZ}.$

\begin{proposition}
At high average SNRs $\left( \text{i.e., }\delta _{SR}=\delta _{RD}=\delta
\rightarrow \infty \right) $, the IP\ of the considered dual-hop UAV-based
WCS\ can be asymptotically expanded, for Scenario II, as
\begin{equation}
P_{int}^{\left( \infty ,2\right) }\sim G_{c}\delta ^{-G_{d}},  \label{ipas}
\end{equation}

where
\begin{equation}
G_{c}=\left\{
\begin{array}{l}
\mathcal{R},m_{SR}N_{R}>m_{RD}N_{D} \\
\mathcal{T}_{1}+\mathcal{T}_{2}-\mathcal{T}_{3},m_{SR}N_{R}<m_{RD}N_{D} \\
\mathcal{R}+\mathcal{T}_{1}+\mathcal{T}_{2}-\mathcal{T}%
_{3},m_{SR}N_{R}=m_{RD}N_{D}%
\end{array}%
\right. ,  \label{Gc}
\end{equation}%
with%
\begin{equation}
\mathcal{R=}\left( \frac{m_{RD}\delta _{RE_{2}}}{m_{RE_{2}}}\right)
^{m_{RD}N_{D}}\frac{\left( m_{RE_{2}}N_{E_{2}}+m_{RD}N_{D}-1\right) !}{%
\left( m_{RE_{2}}N_{E_{2}}-1\right) !\left( m_{RD}N_{D}\right) !},  \label{R}
\end{equation}%
\begin{align}
\mathcal{T}_{1}& =\frac{\left( m_{SR}\gamma _{th}\right)
^{m_{_{SR}}N_{R}}\gamma _{inc}\left( m_{RE_{2}}N_{E_{2}},\frac{m_{RE_{2}}}{%
\delta _{RE_{2}}}\gamma _{th}\right) }{\left( m_{SR}N_{R}\right) !\left(
m_{RE_{2}}N_{E_{2}}-1\right) !}  \notag \\
& \times \frac{\gamma _{inc}\left( m_{SE_{1}}N_{E_{1}},\frac{m_{SE_{1}}}{%
\delta _{SE_{1}}}\gamma _{th}\right) }{\left( m_{SE_{1}}N_{E_{1}}-1\right) !}%
,  \label{T1} \\
\mathcal{T}_{2}& =\frac{m_{SR}^{m_{SR}N_{R}}\left( \mathcal{F}_{SE_{1}}+%
\mathcal{F}_{RE_{2}}\right) }{\left( m_{SR}N_{R}\right) !},  \label{T2} \\
\mathcal{T}_{3}& =\frac{m_{SR}^{m_{SR}N_{R}}\left( \mathcal{G}%
_{S,E_{1},R,E_{2}}+\mathcal{G}_{R,E_{2},S,E_{1}}\right) }{\left(
m_{SR}N_{R}\right) !},  \label{T3}
\end{align}%
\begin{equation}
\mathcal{F}_{x,y}=\frac{\left( \frac{\delta _{xy}}{m_{xy}}\right)
^{m_{SR}N_{R}}\Gamma _{\text{inc}}\left( m_{xy}N_{y}+m_{SR}N_{R},\frac{m_{xy}%
}{\delta _{xy}}\gamma _{th}\right) }{\left( m_{xy}N_{y}-1\right) !},
\end{equation}%
\begin{align}
\mathcal{G}_{x,y,z,t}& =\frac{\left( \frac{m_{xy}}{\delta _{xy}}\right)
^{m_{xy}N_{y}}}{\left( m_{xy}N_{y}-1\right) !}\sum_{n=0}^{m_{zt}N_{t}-1}%
\frac{\left( \frac{m_{zt}}{\delta _{zt}}\right) ^{n}}{n!}  \notag \\
& \times \frac{\Gamma _{\text{inc}}\left( m_{xy}N_{y}+m_{SR}N_{R}+n,\left(
\frac{m_{xy}}{\delta _{xy}}+\frac{m_{zt}}{\delta _{zt}}\right) \gamma
_{th}\right) }{\left( \frac{m_{xy}}{\delta _{xy}}+\frac{m_{zt}}{\delta _{zt}}%
\right) ^{n+m_{xy}N_{y}+m_{SR}N_{R}}},
\end{align}%
and%
\begin{equation}
G_{d}=\min \left( m_{SR}N_{R},m_{RD}N_{D}\right) ,  \label{Gd}
\end{equation}%
are the respective secrecy coding gain and diversity order, respectively.

\begin{proof}
The proof is provided in Appendix B.
\end{proof}
\end{proposition}

\begin{remark}
Equations (\ref{ipas})-(\ref{Gd}) provide an asymptotic expression of the
IP\ in the high SNR\ regime for Scenario II, written in terms of finite
summations, Gamma and lower/upper incomplete Gamma functions, and
exponential terms. Such expression can give insightful observations of the
impact of the setup parameters on the security level in the high SNR\
regime. Interestingly, the achievable secrecy diversity order for this
scenario depends exclusively on the legitimate links' fading parameters and
the number of antennas; i.e., $G_{d}=\min \left(
m_{SR}N_{R},m_{RD}N_{D}\right) $.
\end{remark}

\section{Numerical Results}

In this section, numerical results for the secrecy performance of the
considered dual-hop UAV-based WCS\ are presented. The IP\ metric is
evaluated for several configurations of system and channel parameters.
Unless otherwise stated, the default values of the system parameters are
listed in Table I. Moreover, without loss of generality, the
mobility-dependent noise power can be computed as follows: $\sigma
_{w_{XZ}}^{2}=\Omega _{XZ}+\sigma _{\varepsilon _{XZ}}^{2}$ \cite{anshul}.
It is worth mentioning that throughout the numerical results, the relative
speed between two nodes $X$ and $Z$ can refer to two UAVs moving either in
the same direction, i.e., $v_{XZ}=\left\vert v_{Z}-v_{X}\right\vert $, or in
opposite ones, i.e., $v_{XZ}=v_{X}+v_{Z}$. Also, we use the notation $\delta
=\delta _{SR}=\delta _{RD},$ when the IP is plotted against $\delta _{SR}$, with $%
\delta _{RD}=\delta _{SR}$. Lastly, Monte Carlo simulations were performed
by generating $9\times 10^{6}$ Gamma-distributed RVs, referring to the
per-hop SNRs, with PDF and CDF given by (\ref{pdf}) and (\ref{cdf}),
respectively.
\begin{table*}[t]
\caption{Simulation {p}arameters' {v}alues.}
\label{t1}\centering
\par
\begin{tabular}{c|c}
\hline\hline
\textit{\textbf{Parameter}} & \textit{\textbf{Value}} \\ \hline\hline
$f_{c}$ & $2.4$ GHz \\ \hline
$\tau $ & 1 ms \\ \hline
$v_{X};X\in \left\{ S,R,D,E_{1},E_{2}\right\} $ & 25 Km/h \\ \hline
$m_{XZ};XZ\in \left\{ SR,RD,SE_{1},RE_{2}\right\} $ & 2 \\ \hline
$\delta _{SR}=\delta _{RD}=\delta $ & $30$ dB \\ \hline
$\delta _{SE_{1}}=\delta _{RE_{2}}$ & $10$ dB \\ \hline
$\sigma _{\varepsilon _{XZ}}^{2};XZ\in \left\{ SR,RD,SE_{1},RE_{2}\right\} $
& $0.1$ \\ \hline
$\Omega _{XZ}$ & 2 \\ \hline
$N_{R},N_{D}$ & 4 \\ \hline
$N_{E_{1}},N_{E_{2}}$ & 2 \\ \hline
$\gamma _{th}$ & 3 \\ \hline
\end{tabular}%
\end{table*}

\subsection{Effect of the Fading Severity Parameters and Number of Antennas}

\begin{figure}[tbp]
\begin{center}
\includegraphics[scale=.63]{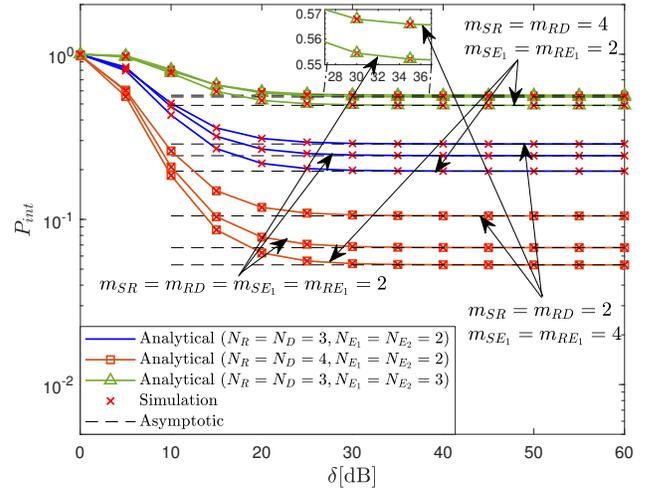}
\end{center}
\caption{IP versus $\protect\delta $ for various $\left( m_{XZ},N_{Z}\right)
_{XZ\in \left\{ SR,RD,SE_{1},RE_{2}\right\} }$ values.}
\label{fig2}
\end{figure}

Fig. \ref{fig2} presents the system's IP\ versus the average legitimate SNR,
i.e., $\delta _{SR}=\delta _{RD}=\delta $, for several combinations of the
fading severity parameter and number of antennas. One can note from this
figure that the analytical curves, plotted using (\ref{ipfinal}), match
their simulations counterpart, which corroborates the accuracy of the
derived IP\ expression. Furthermore, at high SNR, the IP\ justifies with the
asymptotic floors, plotted using (\ref{ipass1})-(\ref{H4}),\ where the\
secrecy diversity order equals zero as pointed out in \textit{Remark 2.2 }and%
\textit{\ Remark 5}. On the other hand, one can note that the system's
secrecy improves by increasing either the legitimate link's fading severity
parameters and/or the number of receive antennas (i.e., higher $m_{SR},$ $%
m_{RD}$ and/or $N_{R},$ $N_{D})$. It is important to note that when the
eavesdropping links are enhanced, either in terms of the number of antennas
onboard and/or the fading severity parameter, the system's secrecy
deteriorates, where the IP\ can reach $50\%\ $\ for $m_{XZ}=2$ $\left( XZ\in
\left\{ SR,\text{ }RD,\text{ }RE_{1},\text{ }RE_{2}\right\} \right) $ and $%
N_{Z}=3.$

\subsection{Effect of\ the CSI\ Imperfection Level}

\begin{figure}[tbp]
\begin{center}
\includegraphics[scale=.63]{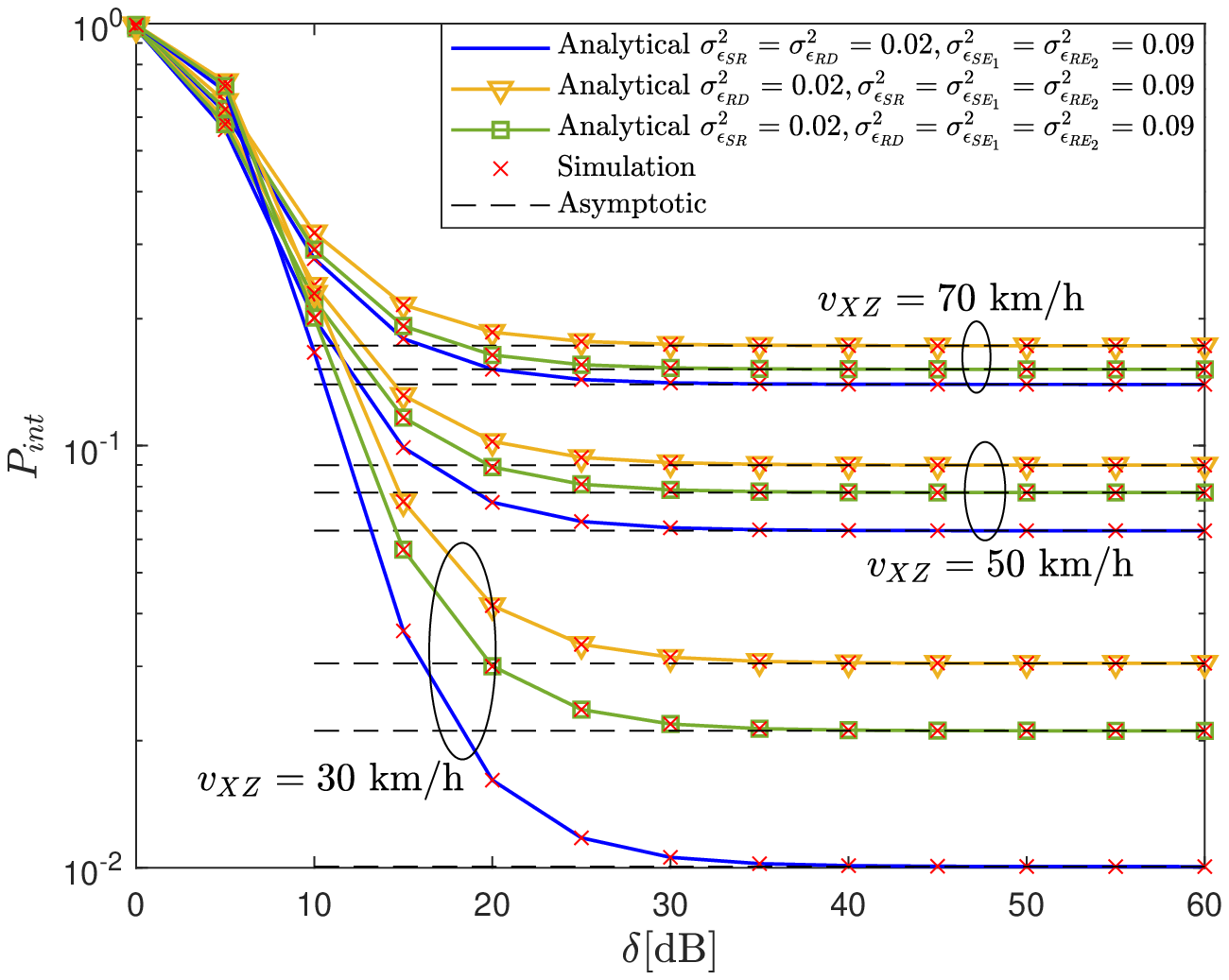}
\end{center}
\caption{IP versus $\protect\delta $ for various $\left( \protect\sigma _{%
\protect\varepsilon _{XZ}}^{2},v_{XZ}\right) _{XZ\in \left\{ SR,\text{ }RD,%
\text{ }SE_{1},\text{ }RE_{2}\right\} }$ values.}
\label{fig3}
\end{figure}

In Fig. \ref{fig3}, the IP\ is depicted vs $\delta _{SR}=\delta _{RD}=\delta
$ for numerous combinations of CSI estimation noise power $\left( \sigma
_{\varepsilon _{XZ}}^{2}\right) $ and nodes relative speed values. We can
notice that CSI\ estimation noise power $\left( \text{i.e., }\sigma
_{\varepsilon _{XZ}}^{2}\right) $ impacts negatively the system's IP.
Specifically, three cases are considered in this figure:\ (i) Case 1:$\ \
\left( \sigma _{\varepsilon _{SR}}^{2}=\sigma _{\varepsilon _{RD}}^{2}=0.02,%
\text{ }\sigma _{\varepsilon _{SE_{1}}}^{2}=\sigma _{\varepsilon
_{RE_{2}}}^{2}=0.09\right) $; i.e., both legitimate links' have less
estimation error noise compared to their wiretap counterpart; (ii) Case 2: $%
\left( \sigma _{\varepsilon _{SR}}^{2}=0.02,\text{ }\sigma _{\varepsilon
_{RD}}^{2}=\sigma _{\varepsilon _{SE_{1}}}^{2}=\sigma _{\varepsilon
_{RE_{2}}}^{2}=0.09\right) $; i.e., estimation error at the $R$-$D$ link
increases; and (iii)\ Case 3:\ $\left( \sigma _{\varepsilon _{RD}}^{2}=0.02,%
\text{ }\sigma _{\varepsilon _{SR}}^{2}=\sigma _{\varepsilon
_{SE_{1}}}^{2}=\sigma _{\varepsilon _{RE_{2}}}^{2}=0.09\right) $; i.e.,
estimation error at the $S$-$R$ link increases. It is obvious that Case 1
exhibits the best secrecy performance when both legitimate links experience
less CSI\ estimation noise compared to their malign counterpart. However, we
can note that Case 2 manifests a slightly improved performance compared to
Case 3. That is, increasing CSI imperfection power on the first legitimate
hop has a more negative impact on the system's secrecy compared to the case
when the CSI imperfection at the second hop is magnified. Lastly, it can be
ascertained as well that, for fixed $\sigma _{\varepsilon _{XZ}}^{2}$, $%
\left( \text{i.e., }\sigma _{\varepsilon _{SR}}^{2}=0.02;\sigma
_{\varepsilon _{RD}}^{2}=\sigma _{\varepsilon _{SE_{1}}}^{2}=\sigma
_{\varepsilon _{RE_{2}}}^{2}=0.09\right) ,$ the system's secrecy degrades
when the nodes' relative speed increases, where it can reach $17.24\%$ at $%
v_{XZ}=70$ km/h.
\begin{figure}[h]
\begin{center}
\includegraphics[scale=0.63]{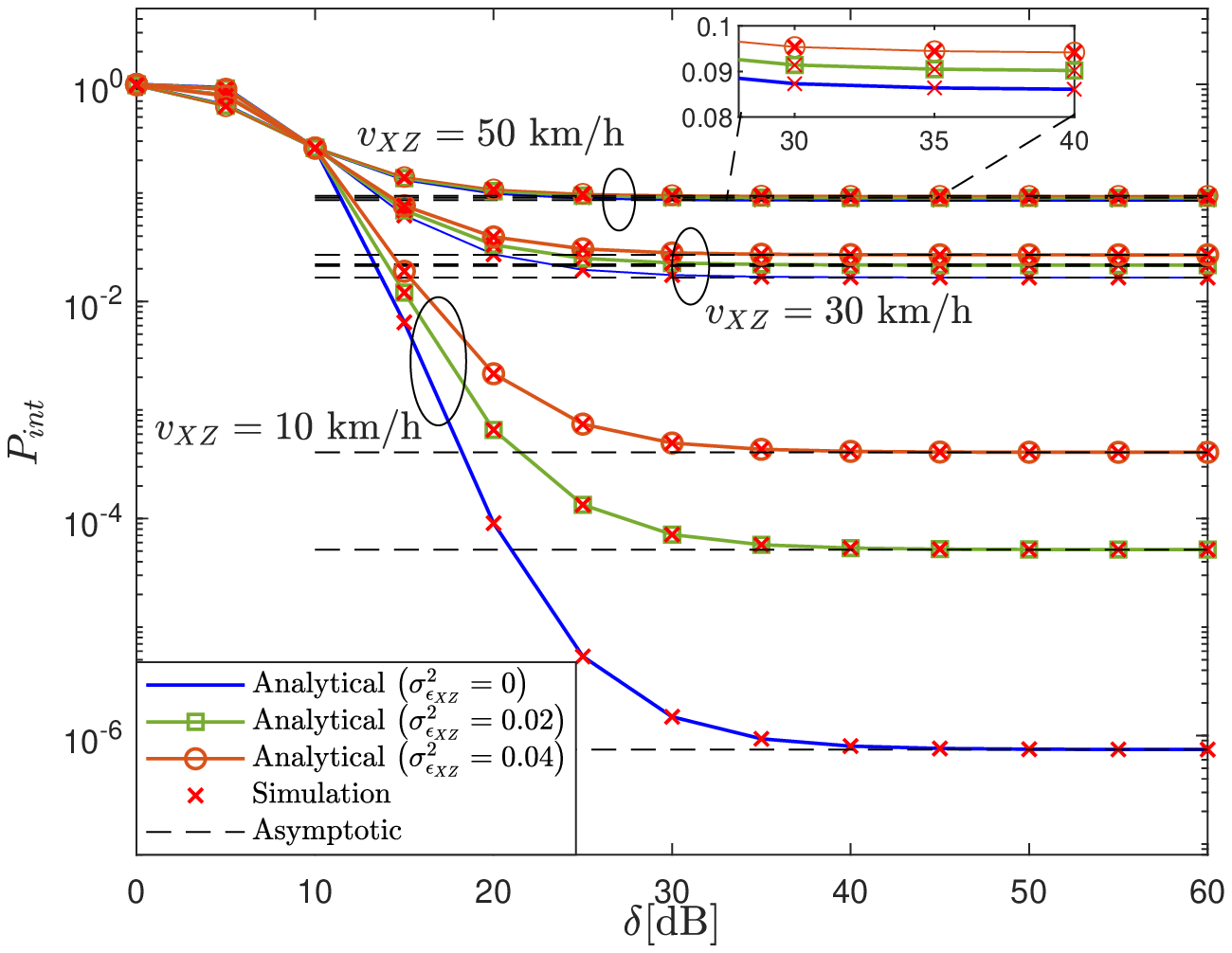}
\end{center}
\caption{IP versus $\protect\delta $ for various $\left( \protect\sigma _{%
\protect\varepsilon _{XZ}}^{2},v_{XZ}\right) _{XZ\in \left\{ SR,\text{ }RD,%
\text{ }SE_{1},\text{ }RE_{2}\right\} }$ values.}
\label{fig4a}
\end{figure}
\begin{figure}[h]
\begin{center}
\includegraphics[scale=0.63]{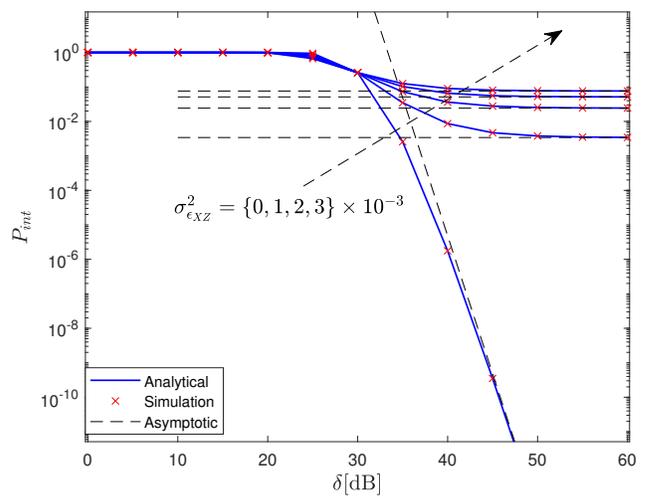}
\end{center}
\caption{IP versus $\protect\delta $ for various $\left( \protect\sigma _{%
\protect\varepsilon _{XZ}}^{2}\right) _{XZ\in \left\{ SR,\text{ }RD,\text{ }%
SE_{1},\text{ }RE_{2}\right\} }$ values and static nodes.}
\label{fig4b}
\end{figure}

To emphasize Fig. \ref{fig3} results, Fig. \ref{fig4a} represent the IP
behavior versus $\delta $ for equal $\sigma _{\varepsilon _{XZ}}^{2}$ on all
links, and various speed values. Notably, for perfect CSI (solid lines)$,$
the IP\ exhibits a remarkable increase by a factor of $10^{4}$ when the
relative speed of nodes increases from 10 to 30 km/h. The degradation,
however, is less significant with respect to the speed, for the remaining
two levels of CSI\ imperfection. Moreover, at 50 km/h, there is almost no
noticeable impact of the CSI\ estimation noise level on the performance.

From another front, Fig. \ref{fig4b} shows the IP\ performance for static
nodes, with $\delta _{SE_{1}}=\delta _{RE_{2}}=30$ dB, and considering
various levels of CSI imperfections $\left( \sigma _{\varepsilon
_{XZ}}^{2}\right) $. The results emphatically highlight the IP\ linear drop
for the ideal case $\left( \text{i.e., }v_{XZ}=\sigma _{\varepsilon
_{XZ}}^{2}=0\right) ,$ where the analytical IP\ curve matches its asymptotic
counterpart, plotted from (\ref{ipas}), at high $\delta $\ values. In such a
case, $\delta \rightarrow \infty \Longleftrightarrow \Upsilon _{SR},$ $%
\Upsilon _{RD}\rightarrow \infty $. Thus, the achievable diversity order for
this case equals $\min \left( m_{SR}N_{R},\text{ }m_{RD}N_{D}\right) $ as
discussed in \textit{Remark 6}; i.e., the achievable diversity order equals $%
8$ for this ideal setup $\left( m_{SR}=m_{RD}=2,\text{ }N_{R}=N_{D}=4\right)
.$ On the other hand, one can note in parallel that even slight increases in
$\sigma _{\varepsilon _{XZ}}^{2}$ can drastically impact the performance,
where the IP\ exhibits a 15 dB loss at $IP\approx 3\times 10^{-3}$ when
going from the ideal case to $\sigma _{\varepsilon _{XZ}}^{2}=10^{-3}.$
Also, one can observe that the IP\ for the imperfect CSI case reaches
asymptotic floors at higher $\delta $ values, where the corresponding
legitimate average SNR, given by (\ref{avgsnrbranch}), reaches a ceiling
value, as detailed in \textit{Remark 2.2}. Thus, the secrecy diversity order
in such a scenario equals zero, as pointed out in \textit{Remark 5}. \

\subsection{Effect of the Legitimate/Wiretap Average SNRs}

\begin{figure}[tbp]
\begin{center}
\includegraphics[scale=.63]{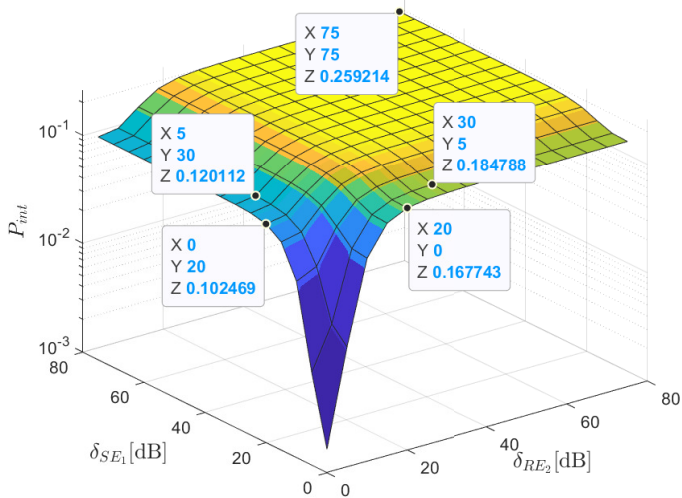}
\end{center}
\caption{IP versus $\protect\delta _{SE_{1}}$ and $\protect\delta _{RE_{2}}$%
. }
\label{fig5}
\end{figure}

Fig. \ref{fig5} provides the IP\ plot in three dimensions versus the average
SNR of the wiretap links, i.e., \ $\delta _{SE_{1}}$ and $\delta _{RE_{2}}$,
with $\delta _{SR}=\delta _{RD}=40$ dB$.$ Expectedly, the IP\ is an
increasing function of both malign channels' SNRs. Notably, at low\ SNR
values, we notice that the IP does not exhibit symmetric behavior for both
eavesdropper's SNRs. Importantly, at low and mid SNR\ values, and $\beta
>\alpha $ $\left( \alpha ,\text{ }\beta \in
\mathbb{R}
_{+}^{\ast }\right) $, the IP\ evaluated at $\delta _{SE_{1}}=\alpha $ and $%
\delta _{RE_{2}}=\beta $ exceeds the one evaluated at $\delta
_{SE_{1}}=\beta $ and $\delta _{RE_{2}}=\alpha $; e.g., IP$=16.77\%$ for $%
\left( \delta _{SE_{1}},\text{ }\delta _{RE_{2}}\right) =\left( 0,20\right) $
dB, while IP$=10.24\%$ for $\left( \delta _{SE_{1}},\text{ }\delta
_{RE_{2}}\right) =\left( 20,0\right) $ dB. Therefore, deteriorating the link
quality of the second malign user can provide a better secrecy performance
than worsening the first one's channel. Lastly, the IP\ manifests a
saturation regime at high $\left( \delta _{SE_{1}},\delta _{RE_{2}}\right) $
values, where the IP\ reaches $26\%$. As detailed in \textit{Remark 2.2},
for fixed $\Upsilon _{SR},\Upsilon _{RD}$ values, the average wiretap SNRs $%
\left( \text{i.e., }\Upsilon _{SE_{1}},\text{ }\Upsilon _{RE_{2}}\right) $
exhibit a ceiling value at higher $\delta _{SE_{1}}$ and $\delta _{RE_{2}}.$
Therefore, the IP\ in this regime depends on the remainder of system
parameters (i.e., legitimate SRNRs values, number of antennas, fading
parameters, correlation coefficients, and CSI imperfection level).

\begin{figure}[tbp]
\begin{center}
\includegraphics[scale=.63]{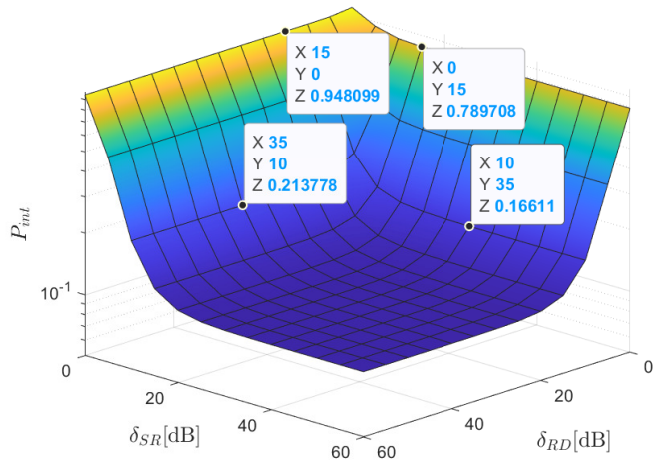}
\end{center}
\caption{IP versus $\protect\delta _{SR}$ and $\protect\delta _{RD}.$}
\label{fig10}
\end{figure}

Fig. \ref{fig10} depicts the secrecy performance in three dimensions versus
the average SNR\ of both hops, i.e., $\delta _{SR}$ and $\delta _{RD}$.
Expectedly, the IP\ is a decreasing function of both $\delta _{SR}$ and $%
\delta _{RD},$ where it reaches a ceiling value as pointed out in \textit{%
Remark 2.2 }and\textit{\ Remark 5}. Also, in a similar fashion to Fig. \ref%
{fig5}, the IP\ does not exhibit symmetric behavior vs $\delta _{SR}$ and $%
\delta _{RD},$ particularly at low SNR\ values. For instance, $%
P_{int}=0.1661 $ for $\left( \delta _{SR},\delta _{RD}\right) =\left(
35,10\right) $ dB, while $P_{int}=0.2138$ for $\left( \delta _{SR},\delta
_{RD}\right) =\left( 10,35\right) $ dB$.$ Thus, the IP\ can be improved by
strengthening the first hop's channel quality.

\subsection{Effect of the Decoding Threshold SNR}

\begin{figure}[tbp]
\begin{center}
\includegraphics[scale=.63]{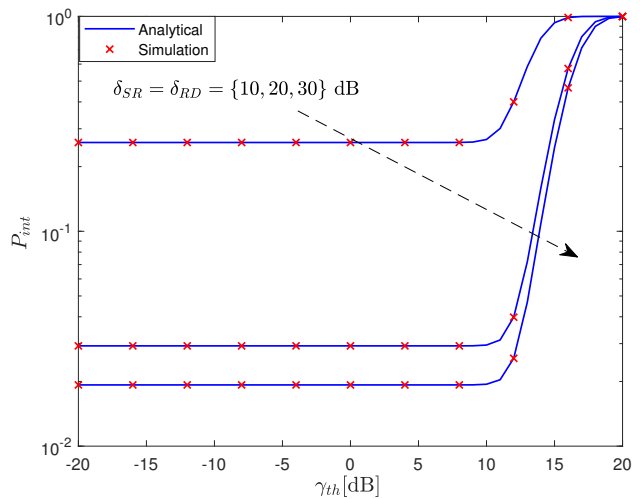}
\end{center}
\caption{IP versus $\protect\gamma _{th}$ for various $\left( \protect\delta %
_{XZ}\right) _{XZ\in \left\{ SR,RD,SE_{1},RE_{2}\right\} }$ values.}
\label{fig6}
\end{figure}

In Fig. \ref{fig6}, the system secrecy performance is shown vs $\gamma _{th}$%
, with $v_{XZ}=20$ km/h $\left( XZ\in \left\{ SR,SE_{1},RD,RE_{2}\right\}
\right) .$ It can be ascertained that the IP\ reaches error floors at low $%
\gamma _{th}$ values, while at higher $\gamma _{th}$, it reaches unit.
Indeed, as discussed in \textit{Remark 4.2}, the terms $\mathcal{I}^{(i)}$
of (\ref{ipfinal}), given by (\ref{I1})-(\ref{I4}), reduce to (\ref{v1})-(%
\ref{v4}), which are independent of $\gamma _{th}$. Furthermore, as
highlighted in \textit{Remark 4.3}, the IP\ was shown to converge to one at
high $\gamma _{th}$, i.e., the higher $\gamma _{th}$, the lower is the
decoding success probability at $R$. Henceforth, this results in a zero
capacity of the second hop, leading to a second hop's secrecy capacity equal
to zero, according to (\ref{cs2}). Thus, relying on the IP\ definition in (%
\ref{ipnew})-(\ref{csmin}), the IP will be evidently equal to 1.

\subsection{Effect of the Carrier Frequency and Delay}

\begin{figure}[tbp]
\begin{center}
\includegraphics[scale=.63]{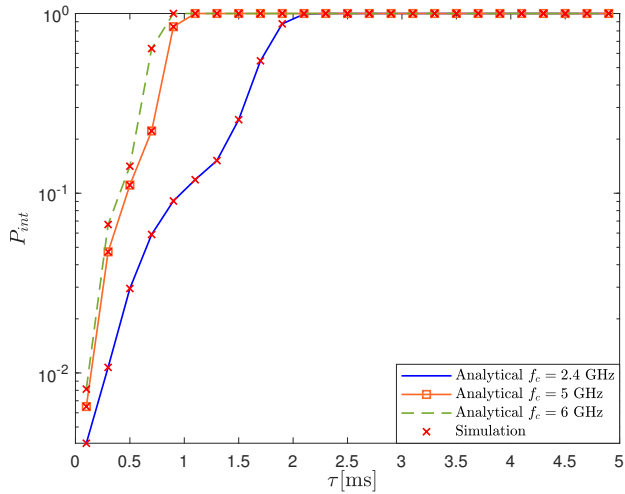}
\end{center}
\caption{IP versus $\protect\tau $ for three main carrier frequency $f_{c}$
values.}
\label{fig7}
\end{figure}

Fig. \ref{fig7} presents the system's IP with respect to the CSI\ estimation
time delay $\tau ,$ for three main carrier frequency values, i.e., $%
f_{c}=2.4 $, $5,$ and $6$ GHz\footnote{%
The considered carrier frequencies in this figures are chosen according to
the Wi-Fi 802.11a, 802.11ac, and 802.11ax standards, with a dynamic range of
channel estimation delay values; i.e., $\tau $. Nevertheless, the evaluation
can be easily extended to other frequency bands (e.g., LTE, mmWave) and
arbitrary delay values.}, with $\delta _{SR}=\delta _{RD}=40$ dB. It is
obvious from the figure that the higher the delay and/or the carrier
frequency, the worse is the IP\ performance. Indeed, as discuss in \textit{%
Remark 1}, $\rho _{XZ}$, given by (\ref{corr}), is decreasing over the
interval $\left[ 0,\frac{2\pi v_{XZ}f_{c}\tau _{0}}{c}=3.8317\right] ,$
where $\tau _{0}\ $equals $5.5,$ $2.6,$ and $2.2$ ms for $f_{c}=2.4,$ $5,$
and $6$ GHz, respectively. Notably, the IP reaches one at $\tau =2.1,$ $1.1,$
and $0.9$ ms, for $f_{c}=2.4,$ $5,$ and $6$ GHz, respectively. Indeed, as
mentioned above, the correlation coefficient decreases over the intervals $%
\left[ 0,5.5\right] $ ms$,$ $\left[ 0,2.6\right] $ ms$,$ and $\left[ 0,2.2%
\right] $ ms for the three considered frequencies in ascending order. Thus,
capitalizing on \textit{Remark 2.1}, it yields a decrease as well in the
per-hop average SNR, given by (\ref{avgsnrbranch}), over these
aforementioned intervals, leading to an increase in the system's IP.
Importantly, we can observe that the IP\ remains constant at one, although
the correlation coefficient reincreases (i.e., reincrease of $\Upsilon _{SR}$
and $\Upsilon _{RD}$) when $\tau _{0}>2.6$ ms and $\tau _{0}>2.2$ ms, for $%
f_{c}=$ $5,$ and $6$ GHz, respectively.

\subsection{Effect of Nodes' Speed}

\begin{figure}[tbp]
\begin{center}
\includegraphics[scale=.63]{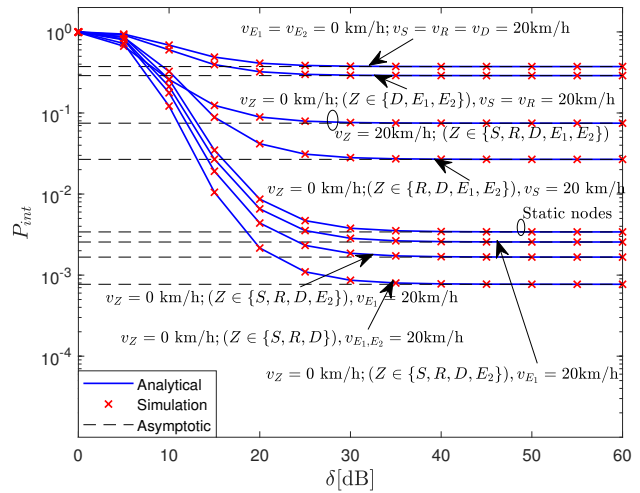}
\end{center}
\caption{IP versus $\protect\delta $ for various values of $\left(
v_{S},v_{R},v_{D},v_{E_{1}},v_{E_{2}}\right) $.}
\label{fig8}
\end{figure}

Fig. \ref{fig8} presents the IP\ behavior vs $\delta $, for various nodes'
mobility scenarios. We assume that $v_{XZ}=v_{X}+v_{Z}$ $\left( XZ\in
\left\{ SR,\text{ }RD,\text{ }SE_{1},\text{ }RE_{2}\right\} \right) ,$ which
corresponds to the case of two moving UAVs in opposite directions. It can be
noted that the worst performance corresponds to the case of moving
legitimate nodes and static wiretap ones. The lesser the number of moving
benign nodes, the better is the performance. Indeed, as emphasized in
\textit{Remark 1}, and for $v_{XZ}\in \left[ 0,\frac{3.8317c}{2\pi f_{c}\tau
_{0}}\right] $, the greater the relative speed $v_{XZ}$, the lower is the
correlation coefficient.\ Thus, the respective average SNR\ decreases, as
shown in \textit{Remark 2.1}. We can note as well that the best performance
corresponds to the case of moving wiretappers and static legitimate nodes.

\begin{figure}[tbp]
\begin{center}
\includegraphics[scale=.63]{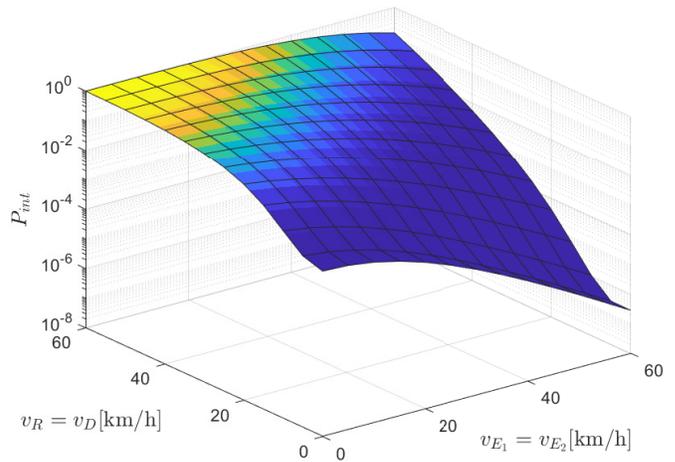}
\end{center}
\caption{IP versus $v_{R}$ and $v_{E_{1}}.$}
\label{fig9}
\end{figure}

Fig. \ref{fig9} shows the IP\ evolution in three dimensions, versus $v_{R}$
and $v_{E_{1}}.$ We set $v_{S}=0$, $v_{D}=v_{R},$ and $v_{E_{2}}=v_{E_{1}}.$
It can be noted that, for fixed legitimate nodes' speed, the IP\ manifests a
decreasing behavior vs the wiretappers' speed. In fact, as discussed in
\textit{Remark 1}, the greater the speed, the lower the correlation
coefficient, leading to a decrease in the average SNR. Likewise, the IP
deteriorates for fixed malign nodes speed and increasing velocity of the
benign ones.

\section{Conclusion}

In this paper, the secrecy performance of a dual-hop UAV-based WCS\ system
was analyzed. Particularly, a DF relay handles the information signal from
a source to a destination, under the presence of two malicious nodes
attempting independently to overhear the $S$-$R$ and $R$-$D$ hops. Transmitters ($S, R$) are assumed to be equipped with a single
transmit antenna, while receivers ($R,D, E_1, E_2$) perform maximal-ratio combining using their multiple receive antennas. Due to
nodes' mobility and channel estimation imperfections at the receivers, the
CSI\ is subject to time selectivity and estimation errors. The per-hop
instantaneous SNRs were expressed in terms of the mobility-dependent
correlation coefficient and CSI\ estimation errors, by using the $AR\left(
1\right) $ model along with Gaussian estimation errors. Using per-hop SNR\ statistics, exact closed-form and asymptotic
expressions for the IP\ metric were derived. In particular, asymptotic
expressions were retrieved for two scenarios: (i) mobile nodes with
imperfect CSI (general case), and (ii) static nodes with perfect
CSI estimation (ideal case). It was shown that the IP admits a zero
diversity order for the first scenario, due to the presence of a ceiling
value of the average SNR\ per branch. However, the IP\ decreases linearly
versus the average SNR in the second one, with an achievable diversity order
dependant on the fading parameters and number of antennas of the legitimate
links/nodes. Furthermore, for static nodes, the system can be castigated by
a $15$ dB\ secrecy loss at IP$=3\times 10^{-3},$ when the CSI imperfection
power raises from $0$ to $10^{-3}$. Finally, we showed that the higher the legitimate nodes'
speed and/or carrier frequency and/or delay time and/or relay's decoding
threshold SNR, the worse is the system's secrecy.

\section*{Appendix A:\ Proof of Proposition 1}

The IP\ of a dual-hop DF-based WCS, subject to the presence of an
eavesdropper per hop is given as \cite[Eq. (33)]{ojcoms}
\begin{equation}
P_{int}=1-\int_{y=\gamma _{th}}^{\infty }f_{\gamma _{SR}}\left( y\right)
F_{\gamma _{SE_{1}}}\left( y\right) \mathcal{J}(y)dy,  \label{propip}
\end{equation}%
where%
\begin{equation}
\mathcal{J}(y)=\int_{z=0}^{y}f_{\gamma _{RE_{2}}}\left( z\right) F_{\gamma
_{RD}}^{\left( c\right) }\left( z\right) dz,  \label{J}
\end{equation}%
and $F_{.}^{\left( c\right) }\left( .\right) =1-F_{.}\left( .\right) $
accounts for the complementary CDF. By plugging the PDF\ and CDF from (\ref%
{pdf}) and (\ref{cdf}), with $XZ=RE_{2}$ and $XZ=RD,$ respectively, into (%
\ref{J}), it yields (\ref{stpaa})-(\ref{stpc}) at the top of the next page,
where Step\textit{\ (a)} holds by involving the expressions (\ref{pdf}) and (%
\ref{cdf}) into (\ref{J}), while Step\textit{\ (b) }is reached by using the
finite-sum representation of $\Gamma _{\text{inc}}\left( .,.\right) $ \cite[%
Eq. (8.352.2)]{integrals} along with \cite[Eq. (8.339.1)]{integrals},
assuming $m_{XZ}\in
\mathbb{N}
^{\ast }$. Finally, using \cite[Eq. (3.381.1)]{integrals}, Step \textit{(c)}
is attained.

By involving (\ref{pdf}) and (\ref{cdf}) with $XZ=SR$ and $XZ=SE_{1},$
respectively, along with (\ref{stpc}) into (\ref{propip}), using \cite[Eq.
(8.339.1)]{integrals} for $m_{XZ}\in
\mathbb{N}
^{\ast }$, and performing some algebraic manipulations, one obtains (\ref%
{pint1}), with $\mathcal{I}$ and $\left( \mathcal{I}^{(i)}\right) _{1\leq
i\leq 4},$ given in (\ref{Is1})-(\ref{I4p}), defined in the next page, where
$\Delta $ and $\Psi $ are defined in the proposition. Step \textit{(a)} is $%
\mathcal{I}$'s definition, Step \textit{(b)} yields from using the
finite-sum representation of $\gamma _{\text{inc}}\left( .,.\right) $ \cite[%
Eq. (8.352.1)]{integrals}, while Step \textit{(c)} and (\ref{I1p})-(\ref{I4p}%
) are obtained by expanding Step \textit{(b)}.\ Finally, armed by \cite[Eq.
(3.381.3)]{integrals} to solve (\ref{I1p})-(\ref{I4p}), (\ref{I1})-(\ref{I4}%
) of \textit{Proposition 1} are obtained. This concludes the proposition's
proof. \qquad
\begin{figure*}[t]
{\normalsize 
\setcounter{mytempeqncnt}{\value{equation}}
\setcounter{equation}{60} }
\par
\begin{align}
\mathcal{J}(y)&\overset{(a)}{=}\frac{\left( \frac{m_{RE_{2}}}{\Upsilon
_{RE_{2}}}\right) ^{m_{RE_{2}}N_{E_{2}}}}{\Gamma \left(
m_{RE_{2}}N_{E_{2}}\right) \Gamma \left( m_{RD}N_{D}\right) }%
\int_{z=0}^{y}z^{m_{RE_{2}}N_{E_{2}}-1}\exp \left( -\frac{m_{RE_{2}}}{%
\Upsilon _{RE_{2}}}z\right) \Gamma _{inc}\left( m_{RD}N_{D},\frac{m_{RD}}{%
\Upsilon _{RD}}z\right) dz.  \label{stpaa} \\
& \overset{(b)}{=}\frac{\left( \frac{m_{RE_{2}}}{\Upsilon _{RE_{2}}}\right)
^{m_{RE_{2}}N_{E_{2}}}}{\left( m_{RE_{2}}N_{E_{2}}-1\right) !}%
\sum_{l=0}^{m_{RD}N_{D}-1}\frac{\left( \frac{m_{RD}}{\Upsilon _{RD}}\right)
^{l}}{l!}\int_{z=0}^{y}z^{m_{RE_{2}}N_{E_{2}}+l-1}\exp \left( -\left[ \frac{%
m_{RE_{2}}}{\Upsilon _{RE_{2}}}+\frac{m_{RD}}{\Upsilon _{RD}}\right]
z\right) dz.  \label{stpbb} \\
& \overset{(c)}{=}\frac{\left( \frac{m_{RE_{2}}}{\Upsilon _{RE_{2}}}\right)
^{m_{RE_{2}}N_{E_{2}}}}{\left( m_{RE_{2}}N_{E_{2}}-1\right) !}%
\sum_{l=0}^{m_{RD}N_{D}-1}\frac{\left( \frac{m_{RD}}{\Upsilon _{RD}}\right)
^{l}\gamma _{\text{inc}}\left( m_{RE_{2}}N_{E_{2}}+l,\left[ \frac{m_{RE_{2}}%
}{\Upsilon _{RE_{2}}}+\frac{m_{RD}}{\Upsilon _{RD}}\right] y\right) }{l!%
\left[ \frac{m_{RE_{2}}}{\Upsilon _{RE_{2}}}+\frac{m_{RD}}{\Upsilon _{RD}}%
\right] ^{l+m_{RE_{2}}N_{E_{2}}}}.  \label{stpc}
\end{align}%
\par
{\normalsize 
\hrulefill 
\vspace*{4pt} }
\end{figure*}
\begin{figure*}[t]
{\normalsize 
\setcounter{mytempeqncnt}{\value{equation}}
\setcounter{equation}{63} }
\par
\begin{align}
P_{int}& =1-\frac{\left( \frac{m_{SR}}{\Upsilon _{SR}}\right)
^{m_{SR}N_{R}}\left( \frac{m_{RE_{2}}}{\Upsilon _{RE_{2}}}\right)
^{m_{RE_{2}}N_{E_{2}}}}{\left( m_{SR}N_{R}-1\right) !\left(
m_{RE_{2}}N_{E_{2}}-1\right) !}\sum_{l=0}^{m_{RD}N_{D}-1}\frac{\left(
m_{RE_{2}}N_{E_{2}}+l-1\right) !\left( \frac{m_{RD}}{\Upsilon _{RD}}\right)
^{l}}{l!\left[ \frac{m_{RE_{2}}}{\Upsilon _{RE_{2}}}+\frac{m_{RD}}{\Upsilon
_{RD}}\right] ^{m_{RE_{2}}N_{E_{2}}+l}}  \notag \\
& \times \underset{\triangleq \mathcal{I}}{\underbrace{\left[ \int_{y=\gamma
_{th}}^{\infty }y^{m_{SR}N_{R}-1}\exp \left( -\frac{m_{SR}}{\Upsilon _{SR}}%
y\right) \frac{\gamma _{\text{inc}}\left( m_{RE_{2}}N_{E_{2}}+l,\left[ \frac{%
m_{RE_{2}}}{\Upsilon _{RE_{2}}}+\frac{m_{RD}}{\Upsilon _{RD}}\right]
y\right) \gamma _{\text{inc}}\left( m_{SE_{1}}N_{E_{1}},\frac{m_{SE_{1}}}{%
\Upsilon _{SE_{1}}}y\right) }{\left( m_{RE_{2}}N_{E_{2}}+l-1\right) !\left(
m_{SE_{1}}N_{E_{1}}-1\right) !}dy\right] }}.  \label{pint1}
\end{align}%
\par
{\normalsize 
\hrulefill 
\vspace*{4pt} }
\end{figure*}
\begin{figure*}[t]
{\normalsize 
\setcounter{mytempeqncnt}{\value{equation}}
\setcounter{equation}{64} }
\par
\begin{align}
\mathcal{I}&\overset{(a)}{=}\int_{y=\gamma _{th}}^{\infty
}y^{m_{SR}N_{R}-1}\exp \left( -\frac{m_{SR}}{\Upsilon _{SR}}y\right) \frac{%
\gamma _{\text{inc}}\left( m_{RE_{2}}N_{E_{2}}+l,\left[ \frac{m_{RE_{2}}}{%
\Upsilon _{RE_{2}}}+\frac{m_{RD}}{\Upsilon _{RD}}\right] y\right) \gamma _{%
\text{inc}}\left( m_{SE_{1}}N_{E_{1}},\frac{m_{SE_{1}}}{\Upsilon _{SE_{1}}}%
y\right) }{\left( m_{RE_{2}}N_{E_{2}}+l-1\right) !\left(
m_{SE_{1}}N_{E_{1}}-1\right) !}dy.  \label{Is1} \\
& \overset{(b)}{=}\int_{y=\gamma _{th}}^{\infty }y^{m_{SR}N_{R}-1}\exp
\left( -\frac{m_{SR}}{\Upsilon _{SR}}y\right) \left[ 1-\exp \left( -\frac{%
m_{SE_{1}}}{\Upsilon _{SE_{1}}}y\right) \sum_{n=0}^{m_{SE_{1}}N_{E_{1}}-1}%
\frac{\left( \frac{m_{SE_{1}}}{\Upsilon _{SE_{1}}}y\right) ^{n}}{n!}\right]
\notag \\
& \times \left[ 1-\exp \left( -\left[ \frac{m_{RE_{2}}}{\Upsilon _{RE_{2}}}+%
\frac{m_{RD}}{\Upsilon _{RD}}\right] y\right)
\sum_{p=0}^{m_{RE_{2}}N_{E_{2}}+l-1}\frac{\left( \left[ \frac{m_{RE_{2}}}{%
\Upsilon _{RE_{2}}}+\frac{m_{RD}}{\Upsilon _{RD}}\right] y\right) ^{p}}{p!}%
\right] dy.  \label{Is2} \\
& \overset{(c)}{=}\sum\limits_{i=1}^{4}\mathcal{I}^{(i)}.  \label{Istp}
\end{align}%
\par
{\normalsize 
\hrulefill 
\vspace*{4pt} }
\end{figure*}
\begin{figure*}[!htb]
{\normalsize 
\setcounter{mytempeqncnt}{\value{equation}}
\setcounter{equation}{67} }
\par
\begin{align}
\mathcal{I}^{(1)}& =\int_{y=\gamma _{th}}^{\infty }y^{m_{SR}N_{R}-1}\exp
\left( -\frac{m_{SR}}{\Upsilon _{SR}}y\right) dy.  \label{I1p} \\
\mathcal{I}^{(2)}& =-\sum_{n=0}^{m_{SE_{1}}N_{E_{1}}-1}\frac{\left( \frac{%
m_{SE_{1}}}{\Upsilon _{SE_{1}}}\right) ^{n}}{n!}\int_{y=\gamma
_{th}}^{\infty }y^{m_{SR}N_{R}+n-1}\exp \left( -\left[ \frac{m_{SR}}{%
\Upsilon _{SR}}+\frac{m_{SE_{1}}}{\Upsilon _{SE_{1}}}\right] y\right) dy.
\label{I2p} \\
\mathcal{I}^{(3)}& =-\sum_{p=0}^{m_{RE_{2}}N_{E_{2}}+l-1}\frac{\left( \frac{%
m_{RE_{2}}}{\Upsilon _{RE_{2}}}+\frac{m_{RD}}{\Upsilon _{RD}}\right) ^{p}}{p!%
}\int_{y=\gamma _{th}}^{\infty }y^{m_{SR}N_{R}+p-1}\exp \left( -\Delta
y\right) dy.  \label{I3p} \\
\mathcal{I}^{(4)}& =\sum_{n=0}^{m_{SE_{1}}N_{E_{1}}-1}\frac{\left( \frac{%
m_{SE_{1}}}{\Upsilon _{SE_{1}}}\right) ^{n}}{n!}%
\sum_{p=0}^{m_{RE_{2}}N_{E_{2}}+l-1}\frac{\left( \frac{m_{RE_{2}}}{\Upsilon
_{RE_{2}}}+\frac{m_{RD}}{\Upsilon _{RD}}\right) ^{p}}{p!}\int_{y=\gamma
_{th}}^{\infty }y^{m_{SR}N_{R}+n+p-1}\exp \left( -\Psi y\right) dy.
\label{I4p}
\end{align}%
\par
{\normalsize 
\hrulefill 
\vspace*{4pt} }
\end{figure*}

\section*{Appendix B:\ Proof of Proposition 2}

In Scenario II, as mentioned before \textit{Proposition 2}, the effective
average SNRs $\Upsilon _{XZ}$ in (\ref{avgsnrbranch}) reduce to $\delta
_{XZ} $ $\left( \text{i.e., }\Upsilon _{XZ}=\delta _{XZ}\right) $. At the
high SNR\ regime $\left( \text{i.e., }\delta _{SR}=\delta _{RD}=\delta
\rightarrow \infty \right) ,$ and relying on the IP\ definition\ in (\ref%
{propip}), we can asymptotically expand it as follows
\begin{equation}
P_{int}^{\left( \infty ,2\right) }=1-\int_{y=\gamma _{th}}^{\infty
}f_{\gamma _{SR}}^{\left( \infty \right) }\left( y\right) F_{\gamma
_{SE_{1}}}\left( y\right) \mathcal{J}^{\left( \infty \right) }(y)dy,
\label{propipas}
\end{equation}%
with
\begin{equation}
\mathcal{J}^{\left( \infty \right) }(y)=\int_{z=0}^{y}f_{\gamma
_{RE_{2}}}\left( z\right) F_{\gamma _{RD}}^{\left( c,\infty \right) }\left(
z\right) dz,  \label{Jasdef}
\end{equation}%
$f_{\gamma _{SR}}^{\left( \infty \right) }\left( y\right) $, and $F_{\gamma
_{RD}}^{\left( c,\infty \right) }\left( z\right) $ being the asymptotic
representations of $\mathcal{J}(y)$,$\ f_{\gamma _{SR}}\left( y\right) $,
and $F_{\gamma _{RD}}^{\left( c\right) }\left( z\right) $ at the high SNR
values, respectively. In such a regime, using the limit of the exponential
function, i.e., $\lim\limits_{x\rightarrow 0}\exp \left( -x\right) =1$,
along with the incomplete Gamma expansion \cite[Eq. (06.06.06.0001.02)]%
{wolfram}, the PDF\ and CDF\ expressions in (\ref{pdf}) and (\ref{cdf}) can
be expanded asymptotically as
\begin{equation}
f_{\gamma _{XZ}}^{\left( \infty \right) }\left( y\right) =\frac{\left( \frac{%
m_{XZ}}{\delta _{XZ}}\right) ^{m_{XZ}N_{Z}}}{\Gamma \left(
m_{XZ}N_{Z}\right) }y^{m_{XZ}N_{Z}-1},  \label{pdfas}
\end{equation}%
and
\begin{equation}
F_{\gamma _{XZ}}^{\left( \infty \right) }\left( y\right) =\frac{\left( \frac{%
m_{XZ}}{\delta _{XZ}}y\right) ^{m_{XZ}N_{Z}}}{\Gamma \left(
m_{XZ}N_{Z}+1\right) },  \label{cdfas}
\end{equation}%
\begin{figure*}[h]
{\normalsize 
\setcounter{mytempeqncnt}{\value{equation}}
\setcounter{equation}{75} }
\par
\begin{align}
\mathcal{J}^{\left( \infty \right) }(y)&\overset{(a)}{=}\frac{\left( \frac{%
m_{RE_{2}}}{\delta _{RE_{2}}}\right) ^{m_{RE_{2}}N_{E_{2}}}}{\left(
m_{RE_{2}}N_{E_{2}}-1\right) !}\int_{z=0}^{y}z^{m_{RE_{2}}N_{E_{2}}-1}\exp
\left( -\frac{m_{RE_{2}}}{\delta _{RE_{2}}}z\right) \left[ 1-\frac{\left(
\frac{m_{RD}}{\delta }z\right) ^{m_{RD}N_{D}}}{\left( m_{RD}N_{D}\right) !}%
\right] dz.  \label{Jas1} \\
& \overset{(b)}{=}\frac{\gamma _{inc}\left( m_{RE_{2}}N_{E_{2}},\frac{%
m_{RE_{2}}}{\delta _{RE_{2}}}y\right) }{\left( m_{RE_{2}}N_{E_{2}}-1\right) !%
}-\frac{\left( \frac{m_{RD}}{\delta }\right) ^{m_{RD}N_{D}}\left( \frac{%
m_{RE_{2}}}{\delta _{RE_{2}}}\right) ^{m_{RE_{2}}N_{E_{2}}}\gamma
_{inc}\left( m_{RE_{2}}N_{E_{2}}+m_{RD}N_{D},\frac{m_{RE_{2}}}{\delta
_{RE_{2}}}y\right) }{\left( \frac{m_{RE_{2}}}{\delta _{RE_{2}}}\right)
^{m_{RE_{2}}N_{E_{2}}+m_{RD}N_{D}}\left( m_{RE_{2}}N_{E_{2}}-1\right)
!\left( m_{RD}N_{D}\right) !}.  \label{Jas}
\end{align}%
{\normalsize 
\hrulefill 
\vspace*{4pt} }
\end{figure*}
respectively. Given that $F_{\gamma _{XZ}}^{\left( c,\infty \right) }\left(
y\right) =1-F_{\gamma _{XZ}}^{\left( \infty \right) }\left( y\right) $, by
plugging (\ref{pdf}) with $\left( XZ=RE_{2},\Upsilon _{RE_{2}}=\delta
_{RE_{2}}\right) $ and (\ref{cdfas}) with $\left( XZ=RD,\delta _{RD}=\delta
\right) $ into (\ref{Jasdef}), respectively, and using \cite[Eq. (8.339.1)]%
{integrals} for $m_{XZ}\in
\mathbb{N}
^{\ast }$, one obtains (\ref{Jas1})-(\ref{Jas}), shown at the top of the
page after next, where Step \textit{(a)} is the definition of $\mathcal{J}%
^{\left( \infty \right) }(y),$ while Step \textit{(b)}\ is attained using
\cite[Eq. (3.381.1)]{integrals}. By involving (\ref{cdf}) with $\left(
XZ=SE_{1},\Upsilon _{SE_{1}}=\delta _{SE_{1}}\right) $, (\ref{pdfas}) with $%
\left( XZ=SR,\delta _{SR}=\delta \right) $, and (\ref{Jas}) into (\ref%
{propipas}), it yields the asymptotic expansion of the IP\ at the high SNR\
as

\begin{align}
P_{int}^{\left( \infty ,2\right) }& \overset{(a)}{\sim }1-\int_{y=\gamma
_{th}}^{\infty }f_{\gamma _{SR}}^{\left( \infty \right) }\left( y\right)
F_{\gamma _{SE_{1}}}\left( y\right) \mathcal{J}^{\left( \infty \right) }(y)dy
\\
& {\overset{(b)}{\sim }}1-\left[ F_{\gamma _{SR}}^{\left( \infty \right)
}\left( y\right) F_{\gamma _{SE_{1}}}\left( y\right) \mathcal{J}^{\left(
\infty \right) }(y)\right] _{\gamma _{th}}^{\infty }  \notag \\
& +\int_{y=\gamma _{th}}^{\infty }F_{\gamma _{SR}}^{\left( \infty \right)
}\left( y\right) \left[ F_{\gamma _{SE_{1}}}\left( y\right) \mathcal{J}%
^{\left( \infty \right) }(y)\right] ^{\prime }dy \\
& {\overset{(c)}{\sim }}\mathcal{U}_{1}+\mathcal{U}_{2}+\mathcal{U}_{3}+%
\mathcal{U}_{4},  \label{ipas1}
\end{align}%
\begin{figure*}[h]
{\normalsize 
\setcounter{mytempeqncnt}{\value{equation}}
\setcounter{equation}{80} }
\par
\begin{align}
\mathcal{U}_{1}& =\frac{\left( \frac{m_{RD}}{\delta }\right)
^{m_{RD}N_{D}}\left( \frac{m_{RE_{2}}}{\delta _{RE_{2}}}\right)
^{m_{RE_{2}}N_{E_{2}}}\left( m_{RE_{2}}N_{E_{2}}+m_{RD}N_{D}-1\right) !}{%
\left( \frac{m_{RE_{2}}}{\delta _{RE_{2}}}\right)
^{m_{RE_{2}}N_{E_{2}}+m_{RD}N_{D}}\left( m_{RE_{2}}N_{E_{2}}-1\right)
!\left( m_{RD}N_{D}\right) !}.  \label{U1} \\
\mathcal{U}_{2}& =\left[ \frac{\gamma _{inc}\left( m_{RE_{2}}N_{E_{2}},\frac{%
m_{RE_{2}}}{\delta _{RE_{2}}}\gamma _{th}\right) }{\left(
m_{RE_{2}}N_{E_{2}}-1\right) !}-\frac{\left( \frac{m_{RD}}{\delta }\right)
^{m_{RD}N_{D}}\left( \frac{m_{RE_{2}}}{\delta _{RE_{2}}}\right)
^{m_{RE_{2}}N_{E_{2}}}\gamma _{inc}\left( m_{RE_{2}}N_{E_{2}}+m_{RD}N_{D},%
\frac{m_{RE_{2}}}{\delta _{RE_{2}}}\gamma _{th}\right) }{\left( \frac{%
m_{RE_{2}}}{\delta _{RE_{2}}}\right)
^{m_{RE_{2}}N_{E_{2}}+m_{RD}N_{D}}\left( m_{RE_{2}}N_{E_{2}}-1\right)
!\left( m_{RD}N_{D}\right) !}\right]  \notag \\
&\times \frac{\gamma _{inc}\left( m_{SE_{1}}N_{E_{1}},\frac{m_{SE_{1}}}{%
\delta _{SE_{1}}}\gamma _{th}\right) \left( \frac{m_{SR}}{\delta }\gamma
_{th}\right) ^{m_{_{SR}}N_{R}}}{\left( m_{SE_{1}}N_{E_{1}}-1\right) !\left(
m_{SR}N_{R}\right) !}.  \label{U2}
\end{align}%
\par
{\normalsize 
\hrulefill 
\vspace*{4pt} }
\end{figure*}
\begin{figure*}[!htb]
{\normalsize 
\setcounter{mytempeqncnt}{\value{equation}}
\setcounter{equation}{82} }
\par
\begin{align}
\mathcal{U}_{3}& \overset{(a)}{=}\int_{y=\gamma _{th}}^{\infty }F_{\gamma
_{SR}}^{\left( \infty \right) }\left( y\right) f_{\gamma _{SE_{1}}}\left(
y\right) \mathcal{J}^{\left( \infty \right) }(y)dy.  \label{U31} \\
& \overset{(b)}{=}\frac{\left( \frac{m_{SE_{1}}}{\delta _{SE_{1}}}\right)
^{m_{SE_{1}}N_{E_{1}}}\left( \frac{m_{SR}}{\delta }\right) ^{m_{SR}N_{R}}}{%
\left( m_{SR}N_{R}\right) !\left( m_{SE_{1}}N_{E_{1}}-1\right) !}%
\int_{y=\gamma _{th}}^{\infty }y^{m_{SE_{1}}N_{E_{1}}+m_{SR}N_{R}-1}\exp
\left( -\frac{m_{SE_{1}}}{\delta _{SE_{1}}}z\right)  \notag \\
& \times \left[ \frac{\gamma _{inc}\left( m_{RE_{2}}N_{E_{2}},\frac{%
m_{RE_{2}}}{\delta _{RE_{2}}}y\right) }{\left( m_{RE_{2}}N_{E_{2}}-1\right) !%
}-\frac{\left( \frac{m_{RD}}{\delta }\right) ^{m_{RD}N_{D}}\left( \frac{%
m_{RE_{2}}}{\delta _{RE_{2}}}\right) ^{m_{RE_{2}}N_{E_{2}}}}{\left( \frac{%
m_{RE_{2}}}{\delta _{RE_{2}}}\right) ^{m_{RE_{2}}N_{E_{2}}+m_{RD}N_{D}}}%
\frac{\gamma _{inc}\left( m_{RE_{2}}N_{E_{2}}+m_{RD}N_{D},\frac{m_{RE_{2}}}{%
\delta _{RE_{2}}}y\right) }{\left( m_{RE_{2}}N_{E_{2}}-1\right) !\left(
m_{RD}N_{D}\right) !}\right] dy.  \label{U3}
\end{align}%
\par
{\normalsize 
\hrulefill 
\vspace*{4pt} }
\end{figure*}
\begin{figure*}[!htb]
{\normalsize 
\setcounter{mytempeqncnt}{\value{equation}}
\setcounter{equation}{84} }
\par
\begin{align}
\mathcal{U}_{4}& \overset{(a)}{=}\int_{y=\gamma _{th}}^{\infty }F_{\gamma
_{SR}}^{\left( \infty \right) }\left( y\right) F_{\gamma _{SE_{1}}}\left(
y\right) \left[ \mathcal{J}^{\left( \infty \right) }(y)\right] ^{\prime }dy.
\label{U41} \\
& \overset{(b)}{=}\frac{\left( \frac{m_{SR}}{\delta }\right) ^{m_{SR}N_{R}}}{%
\left( m_{SR}N_{R}\right) !\left( m_{SE_{1}}N_{E_{1}}-1\right) !}%
\int_{y=\gamma _{th}}^{\infty }y^{m_{SR}N_{R}}\gamma _{inc}\left(
m_{SE_{1}}N_{E_{1}},\frac{m_{SE_{1}}}{\delta _{SE_{1}}}y\right)  \notag \\
& \times \left[
\begin{array}{c}
\frac{\exp \left( -\frac{m_{RE_{2}}}{\delta _{RE_{2}}}y\right) \left( \frac{%
m_{RE_{2}}}{\delta _{RE_{2}}}\right)
^{m_{RE_{2}}N_{E_{2}}}y^{m_{RE_{2}}N_{E_{2}}-1}}{\left(
m_{RE_{2}}N_{E_{2}}-1\right) !}-\frac{\left( \frac{m_{RD}}{\delta }\right)
^{m_{RD}N_{D}}\left( \frac{m_{RE_{2}}}{\delta _{RE_{2}}}\right)
^{m_{RE_{2}}N_{E_{2}}}}{\left( \frac{m_{RE_{2}}}{\delta _{RE_{2}}}\right)
^{m_{RE_{2}}N_{E_{2}}+m_{RD}N_{D}}} \\
\times \frac{\exp \left( -\frac{m_{RE_{2}}}{\delta _{RE_{2}}}y\right) \left(
\frac{m_{RE_{2}}}{\delta _{RE_{2}}}\right)
^{m_{RE_{2}}N_{E_{2}}+m_{RD}N_{D}}y^{m_{RE_{2}}N_{E_{2}}+m_{RD}N_{D}-1}}{%
\left( m_{RE_{2}}N_{E_{2}}-1\right) !\left( m_{RD}N_{D}\right) !}%
\end{array}%
\right] dy.  \label{U4}
\end{align}%
\par
{\normalsize 
\hrulefill 
\vspace*{4pt} }
\end{figure*}
where Step\textit{\ (a)}\ is the definition of IP\ at high SNR, Step \textit{%
(b)}\ holds by using integration by parts with $u^{\prime }=f_{\gamma
_{SR}}^{\left( \infty \right) }\left( y\right) $ and $v=F_{\gamma
_{SE_{1}}}\left( y\right) \mathcal{J}^{\left( \infty \right) }(y),$ and Step%
\textit{\ (c)} is reached after using \cite[Eq. (8.339.1)]{integrals} for $%
m_{XZ}\in
\mathbb{N}
^{\ast }$ alongside some algebraic manipulations and simplifications, where $%
\left( \mathcal{U}_{i}\right) _{1\leq i\leq 4}$ are given by (\ref{U1})-(\ref%
{U4}). Step \textit{(b)} of $\mathcal{U}_{3}$ in (\ref{U3}) yields from
involving (\ref{cdfas}) and (\ref{pdf}), with $XZ=SR$ and $XZ=SE_{1}$,
respectively, along with (\ref{J}) into the corresponding prior step. On the
other hand, Step \textit{(b)} of $\mathcal{U}_{4}$ in (\ref{U4}) is produced
by plugging into its initial step (\ref{cdfas}) and (\ref{cdf}), with $%
\left( XZ=SR,\delta _{SR}=\delta \right) $ and $\left( XZ=SE_{1},\Upsilon
_{SE_{1}}=\delta _{SE_{1}}\right) $, respectively, jointly with the first
derivative of $\mathcal{J}^{\left( \infty \right) }(y)$ with respect to $y$,
computed using the derivative of the incomplete Gamma function \cite[Eq.
(06.06.20.0003.01)]{wolfram}.

Thus, one can note from (\ref{ipas1}) that the IP's asymptotic expression is
the sum of four terms, given by (\ref{U1})-(\ref{U4}), where each term has
either one or two subterms with two different coding gain and diversity
order (i.e., power of $\delta $) pairs, where this latter equals $%
m_{SR}N_{R} $, or $m_{RD}N_{D},$ or $m_{SR}N_{R}+m_{RD}N_{D}$ in these
terms/subterms. As a result, only the terms with the least diversity order
are summed up together. Hence, we distinguish three cases:

\subsection{Case I:\ $m_{SR}N_{R}>m_{RD}N_{D}$}

In such a case, only the terms/subterms from (\ref{U1})-(\ref{U4}), for
which the diversity order equals $m_{RD}N_{D}$ are considered. As a result,
this yields taking into account only (\ref{U1}), which corresponds to the
first case of (\ref{Gc}).

\subsection{Case II:\ $m_{SR}N_{R}<m_{RD}N_{D}$}

In this scenario, we take into account only the terms/subterms from (\ref{U1}%
)-(\ref{U4}) that exhibit a diversity order of $m_{SR}N_{R}$. Hence, it
yields (\ref{case2stpa}) and (\ref{case2stpb}), where Step \textit{(b)} is
reached by using the finite sum representation of the lower-incomplete Gamma
function \cite[Eq. (8.352.1)]{integrals} in Step \textit{(a)}. Finally, by
making use of \cite[Eq. (3.381.3)]{integrals} along with some algebraic
manipulations, the second case of (\ref{Gc}), is attained.
\begin{figure*}[!htb]
{\normalsize 
\setcounter{mytempeqncnt}{\value{equation}}
\setcounter{equation}{86} }
\par
\begin{align}
P_{int}^{\left( \infty \right) }& \overset{(a)}{\sim }\frac{\left( \frac{%
m_{SR}}{\delta }\right) ^{m_{SR}N_{R}}\left( \frac{m_{SE_{1}}}{\delta
_{SE_{1}}}\right) ^{m_{SE_{1}}N_{E_{1}}}}{\left( m_{SR}N_{R}\right) !\left(
m_{SE_{1}}N_{E_{1}}-1\right) !}\int_{y=\gamma _{th}}^{\infty }\left[
y^{m_{SE_{1}}N_{E_{1}}+m_{SR}N_{R}-1}\exp \left( -\frac{m_{SE_{1}}}{\delta
_{SE_{1}}}y\right) \frac{\gamma _{inc}\left( m_{RE_{2}}N_{E_{2}},\frac{%
m_{RE_{2}}}{\delta _{RE_{2}}}y\right) }{\left( m_{RE_{2}}N_{E_{2}}-1\right) !%
}\right] dy  \notag \\
& +\frac{\left( \frac{m_{SR}}{\delta }\right) ^{m_{SR}N_{R}}\left( \frac{%
m_{RE_{2}}}{\delta _{RE_{2}}}\right) ^{m_{RE_{2}}N_{E_{2}}}}{\left(
m_{SR}N_{R}\right) !\left( m_{RE_{2}}N_{E_{2}}-1\right) !}\int_{y=\gamma
_{th}}^{\infty }\left[ y^{m_{SR}N_{R}+m_{RE_{2}}N_{E_{2}}-1}\exp \left( -%
\frac{m_{RE_{2}}}{\delta _{RE_{2}}}y\right) \frac{\gamma _{inc}\left(
m_{SE_{1}}N_{E_{1}},\frac{m_{SE_{1}}}{\delta _{SE_{1}}}y\right) }{\left(
m_{SE_{1}}N_{E_{1}}-1\right) !}\right] dy  \notag \\
& +\frac{\left( \frac{m_{SR}}{\delta }\gamma _{th}\right)
^{m_{_{SR}}N_{R}}\gamma _{inc}\left( m_{RE_{2}}N_{E_{2}},\frac{m_{RE_{2}}}{%
\delta _{RE_{2}}}\gamma _{th}\right) \gamma _{inc}\left( m_{SE_{1}}N_{E_{1}},%
\frac{m_{SE_{1}}}{\delta _{SE_{1}}}\gamma _{th}\right) }{\left(
m_{SR}N_{R}\right) !\left( m_{RE_{2}}N_{E_{2}}-1\right) !\left(
m_{SE_{1}}N_{E_{1}}-1\right) !}.  \label{case2stpa} \\
& \overset{(b)}{\sim }\frac{\left( \frac{m_{SR}}{\delta }\right)
^{m_{SR}N_{R}}\left( \frac{m_{SE_{1}}}{\delta _{SE_{1}}}\right)
^{m_{SE_{1}}N_{E_{1}}}}{\left( m_{SR}N_{R}\right) !\left(
m_{SE_{1}}N_{E_{1}}-1\right) !}\int_{y=\gamma _{th}}^{\infty }\left[
\begin{array}{l}
\left( 1-\exp \left( -\frac{m_{RE_{2}}}{\delta _{RE_{2}}}y\right)
\sum\limits_{n=0}^{m_{RE_{2}}N_{E_{2}}-1}\frac{\left( \frac{m_{RE_{2}}}{%
\delta _{RE_{2}}}y\right) ^{n}}{n!}\right) \\
\times y^{m_{SE_{1}}N_{E_{1}}+m_{SR}N_{R}-1}\exp \left( -\frac{m_{SE_{1}}}{%
\delta _{SE_{1}}}y\right)%
\end{array}%
\right] dy  \notag \\
& +\frac{\left( \frac{m_{SR}}{\delta }\right) ^{m_{SR}N_{R}}\left( \frac{%
m_{RE_{2}}}{\delta _{RE_{2}}}\right) ^{m_{RE_{2}}N_{E_{2}}}}{\left(
m_{SR}N_{R}\right) !\left( m_{RE_{2}}N_{E_{2}}-1\right) !}\int_{y=\gamma
_{th}}^{\infty }\left[
\begin{array}{l}
\left( 1-\exp \left( -\frac{m_{SE_{1}}}{\delta _{SE_{1}}}y\right)
\sum\limits_{n=0}^{m_{SE_{1}}N_{E_{1}}-1}\frac{\left( \frac{m_{SE_{1}}}{%
\delta _{SE_{1}}}y\right) ^{n}}{n!}\right) \\
\times y^{m_{SR}N_{R}+m_{RE_{2}}N_{E_{2}}-1}\exp \left( -\frac{m_{RE_{2}}}{%
\delta _{RE_{2}}}y\right)%
\end{array}%
\right] dy  \notag \\
& +\frac{\left( \frac{m_{SR}}{\delta }\gamma _{th}\right)
^{m_{_{SR}}N_{R}}\gamma _{inc}\left( m_{RE_{2}}N_{E_{2}},\frac{m_{RE_{2}}}{%
\delta _{RE_{2}}}\gamma _{th}\right) \gamma _{inc}\left( m_{SE_{1}}N_{E_{1}},%
\frac{m_{SE_{1}}}{\delta _{SE_{1}}}\gamma _{th}\right) }{\left(
m_{SR}N_{R}\right) !\left( m_{RE_{2}}N_{E_{2}}-1\right) !\left(
m_{SE_{1}}N_{E_{1}}-1\right) !}.  \label{case2stpb}
\end{align}%
\par
{\normalsize 
\hrulefill 
\vspace*{4pt} }
\end{figure*}

\subsection{Case III:\ $m_{SR}N_{R}=m_{RD}N_{D}$}

In this case, the terms considered in the two previous cases will be summed
up, producing the third case of (\ref{Gc}). This concludes the proof of
\textit{Proposition 2}. \qquad

\bibliographystyle{IEEEtran}
\bibliography{refs}


\end{document}